\documentclass[12pt]{article}
\usepackage{graphicx}
\usepackage{amssymb}
\usepackage{amsmath}
\usepackage{subfigure}
\usepackage{cite}
\usepackage{ulem}
\input{colordvi.tex}

\usepackage{epsfig,amssymb,amsmath,color}
\usepackage[colorlinks=true,citecolor=blue,linkcolor=blue]{hyperref}

\begin{document}

\def\beq{\begin{equation}}
\def\eeq{\end{equation}}

\def\bea{\begin{eqnarray}}
\def\eea{\end{eqnarray}}

\def\bi{\begin{itemize}}
\def\ei{\end{itemize}}

\def\fr{{\rm fr}}
\def\osc{{\rm osc}}
\def\dec{{\rm dec}}
\def\pl{{\rm pl}}
\def\f{{\cal F}}

\begin{titlepage}

\begin{flushright}

RUP-17-11
\end{flushright}

\vskip 1.35cm

\begin{center}

{\Large 
{\bf Refined Study of Isocurvature Fluctuations \\
in the Curvaton Scenario} 
}

\vskip 1.2cm

Naoya Kitajima$^{a,b}$,
David Langlois$^c$,
Tomo Takahashi$^d$,
and
Shuichiro Yokoyama$^{e,f}$ \\

\vskip 0.4cm

{\it$^a$Department of Physics, Nagoya University, Nagoya 464-8602, Japan}\\
{\it$^b$Asia Pacific Center for Theoretical Physics, Pohang 37673, Korea}\\
{\it$^c$APC (Astroparticules et Cosmologie), CNRS-Universit\'e Paris Diderot, 10, rue Alice Domon et L\'eonie Duquet, 75013 Paris, France}\\
{\it $^d$Department of Physics, Saga University, Saga 840-8502, Japan }\\
{\it $^e$Department of Physics, Rikkyo University, 3-34-1 Nishi-Ikebukuro, Toshima, Tokyo, 171-8501, Japan}\\
{\it $^f$Kavli IPMU (WPI), UTIAS, The University of Tokyo,
Kashiwa, Chiba 277-8583, Japan}\\

\date{\today}

\begin{abstract} 
We revisit  the generation of  dark matter isocurvature perturbations in the curvaton model in greater detail, both analytically and numerically.
As concrete examples, we investigate the
cases of 
 thermally decoupled dark matter and axionic dark matter.
We show that the radiation produced by the decay of the curvaton, which has not been taken into account in previous 
analytical studies, can significantly affect  
the amplitude  of isocurvature perturbations. In particular, we find that  they are  drastically suppressed even when the dark matter freeze-out (or the onset of the axion oscillations for axionic dark matter) occurs before the curvaton decays, 
provided the freeze-out takes place deep in the curvaton-dominated Universe.
As a consequence,  we  show that the current observational isocurvature constraints on the curvaton parameters are not as severe as usually thought.

\end{abstract}

\end{center}
\end{titlepage}

\section{Introduction}
\label{sec:intro}

Current cosmological data such as Planck  \cite{Ade:2015xua,Ade:2015lrj} have reached a high level of precision, thereby providing
 detailed information on the nature of primordial density fluctuations. 
Although the fluctuations of the inflaton are usually assumed to be directly at the origin of the observed density fluctuations,  other possibilities can also been envisaged within the framework of inflation. 
 For example, a scalar field other than the inflaton, whose energy density is negligible during 
inflation, could also generate primordial fluctuations,  as in the curvaton scenario \cite{Enqvist:2001zp,Lyth:2001nq,Moroi:2001ct}, which has been widely discussed in the literature.
The current severe upper bound on the tensor-to-scalar $r$
ratio\footnote{ 
From the combination of BICEP2/Keck Array, Planck and other data, the constraint on $r$ is given as $ r < 0.07$ (95~\% C.L.) \cite{Array:2015xqh}.
} is compatible with this type of multi-field models, where  $r$ tends to be small.

In multi-field models, isocurvature fluctuations can also be  produced in addition to  the curvature perturbation. 
Although current observations have severely constrained the amplitude of isocurvature 
fluctuations \cite{Ade:2015lrj}, a small  contribution is still allowed.
Since isocurvature fluctuations can be associated with dark matter (DM) or baryons, a nonzero isocurvature perturbation  would give precious information concerning  models of DM or baryogenesis. 
In the curvaton model, the size of isocurvature fluctuations depends on when and how dark matter particles 
were created and/or the baryon asymmetry of the Universe was generated \cite{Moroi:2002rd,Lyth:2002my,Moroi:2008nn,Takahashi:2009cx,Lyth:2003ip,Lemoine:2006sc}.
Therefore, even if isocurvature fluctuations are not observed,  stringent upper bounds 
 are also  very useful to constrain models of DM and baryogenesis  as they imply  some restrictions  on the generation mechanism of primordial fluctuations.  In  light of these considerations and 
given the current severe constraints,  it is worth revisiting the issue of  
isocurvature fluctuations in the curvaton scenario in greater detail.

As a concrete example, we consider  weakly interacting massive particles (WIMP)   in  curvaton models.
In most previous studies, it has been considered that the size of isocurvature fluctuations is  too large,
given the present observational bounds, in models where WIMP dark matter freezes out {\it before} the curvaton decay, thus excluding such models.   
In the present work, we reexamine this standard lore by investigating the generation of isocurvature fluctuations in detail
and  show that, even if the freeze-out occurs before the curvaton decay,  isocurvature fluctuations  are significantly suppressed  
provided WIMPs freeze out deep in the curvaton-dominated era.
In the curvaton scenario,  the current upper limits  on non-Gaussianity indicate that  the curvaton should dominate the Universe at the time of its decay, which allows a DM freeze-out  deep in the curvaton-dominated epoch. Hence our findings suggest that the non-detection of isocurvature fluctuations may not be as restrictive for curvaton models as previously thought. 

In this article, we also investigate  axionic DM in the curvaton scenario. In this case, the beginning of the axion oscillations is considered to be the critical time that determines the size of isocurvature fluctuations and  it has been argued that, when the axion starts to oscillate before the curvaton decay, 
large (correlated) isocurvature fluctuations\footnote{ 
Here, we have neglected the uncorrelated isocurvature perturbations of axion DM, which could be generated from the quantum fluctuations of the axion field.
} are generated. However, similarly to the WIMP case, 
we  show that, even when the onset of  axion oscillations takes place before the curvaton decay, 
isocurvature fluctuations are drastically suppressed if it occurs  deep in the curvaton-dominated epoch.

The organization of this paper is as follows. In the next section, 
we derive a general formula for  super-Hubble density perturbations in a cosmological scenario where some transition (decay, freeze-out, ...) occurs in the early Universe.
In this analytical derivation, we employ the usual approximation that the transition occurs instantaneously.
In section~\ref{sec:iso_DM}, based on the derived formula we discuss DM isocurvature perturbations in the curvaton scenario for thermally decoupled DM such as WIMP.
We also  use of a numerical analysis and check the validity of the sudden transition approximation.
As another example, we  consider  axionic DM, taking into   account the temperature dependence for the axion mass.
In section~\ref{sec:implications}, we discuss the implications of our analysis for the isocurvature 
constraints on model parameters both in the WIMP and axion cases.
The final section is devoted to the summary of our results.

\section{Sudden transition formalism}
\label{sec:formalism}

Before investigating the DM isocurvature perturbations in the curvaton model, 
 we present in this section a general formalism that enables us 
 to  determine  analytically the  density perturbations  in a cosmological scenario where some transition (decay,  freeze-out, \dots)  occurs in the early Universe. At the time of the transition, we  consider only super-Hubble perturbations, which will later reenter the Hubble radius, and resort to  the  $\delta N$ formalism
 \cite{Starobinsky:1986fxa,Salopek:1990jq,Sasaki:1995aw,Sasaki:1998ug}. 
 In order to  analytically follow the evolution of the perturbations during the transition,  we adopt the so-called sudden transition approximation where the transition  is supposed to occur instantaneously.

\subsection{General formula}
\label{sec:general}

Even if  the curvaton is a scalar field, it can be treated as a fluid  once it starts to oscillate at the bottom of its potential. The corresponding equation of state depends on the shape of the potential.  For example, 
an oscillating scalar field in a quadratic potential behaves as pressureless matter (i.e., its equation of state is $w=0$), which we will assume from now on.
Considering that all matter components are  described as fluids, we can define for each individual fluid the nonlinear  perturbation
\begin{equation}
\label{eq:zeta_i}
\zeta_i = \delta N + \frac{1}{3(1+w_i)} \ln \left( \frac{\rho_i ( t, \vec{x})}{\bar{\rho}_i (t)} \right),
\end{equation}
where $\delta N$ denotes the local perturbation of the number of $e$-folds, 
$\rho_i (t, \vec{x})$  is the energy density of the $i$-th component and $\bar{\rho}_i (t)$ its background  homogeneous value; 
$w_i = \bar{P}_i / \bar{\rho}_i$  is the equation of state parameter for the $i$-th fluid,  $\bar{P}_i$ being the background pressure. 
The definition of $\zeta_i$ in Eq.~\eqref{eq:zeta_i}  can be  inverted to give  
\begin{equation}
\rho_i (t, \vec{x}) 
= \bar{\rho}_i e^{3 (1+w_i) (\zeta_i - \delta N)}.
\end{equation}

We now wish to study the evolution of the matter perturbations through  a cosmological transition, which we assume to be instantaneous on the  transition hypersuface ${\cal H}$  defined by 
\begin{equation}
\label{eq:H_Gamma}
H_{\cal H} = \Gamma_{\cal H},
\end{equation}
where $H$ is the Hubble parameter and $\Gamma$ is a physical quantity that depends on
 the nature of the transition: for a transition due to some particle decay, $\Gamma$ corresponds to the 
decay rate; for a freeze-out transition, $\Gamma$ represents  the interaction rate. 

By combining Eq.~\eqref{eq:H_Gamma} and the (first) Friedmann equation, we can relate the (perturbed) energy density and the transition rate 
as
 \begin{equation}
\label{eq:transition_H}
\frac{\rho_{\rm tot} (t_{\cal H}, \vec{x})}{\bar{\rho}_{\rm tot} (t_{\cal H}) }  
= \frac{H^2  (t_{\cal H}, \vec{x})}{\bar{H}^2  (t_{\cal H})} 
= \frac{\Gamma^2 (t_{\cal H}, \vec{x})}{\bar{\Gamma}^2  (t_{\cal H})}\,,
\end{equation}
where a bar denotes a background homogeneous quantity. 
Inserting in Eq.~\eqref{eq:transition_H}  the expression of  the total energy density in terms of the individual ones,
\begin{equation}
\rho_{\rm tot} (t_{\cal H}, \vec{x}) = \sum_i \rho_i (t_{\cal H}, \vec{x}) = \sum_i \bar{\rho}_i (t_{\cal H}) e^{ 3(1+w_i)  (\zeta_i - \delta N_{\cal H})}\,,
\end{equation}
we obtain
\begin{equation}
\label{eq:Omega_Gamma}
\sum_i \Omega_{i} (t_{\cal H}) e^{ 3(1+w_i) (\zeta_{i} - \delta N_{\cal H})} 
=
\frac{\Gamma^2 (t_{\cal H}, \vec{x})}{\bar{\Gamma}^2  (t_{\cal H})},
\end{equation}
where we have introduced the background parameters $\Omega_i= \bar{\rho}_i (t)/(3M_{\rm pl}^2H^2)$.

In the following, we will restrict our  analysis to linear perturbations, although we stress that the above formalism can easily be extended to nonlinear perturbations (see e.g. \cite{Langlois:2013dh, Langlois:2011zz}). 
Expanding Eq.~\eqref{eq:Omega_Gamma} at linear order, we find that the transition hypersurface is characterized by the  $e$-folding number perturbation 
\begin{equation}
\label{eq:deltaN_general}
\delta N_{\cal H} = - \frac{2}{3 \tilde{\Omega}} \delta_\Gamma + \frac{1}{\tilde{\Omega}} \sum_i \tilde{\Omega}_{i} \zeta_{i}\,,
\end{equation}
where we have introduced  the relative perturbation of the transition rate
\begin{equation}
\delta_\Gamma \equiv \frac{\Gamma (t_{\cal H}, \vec{x})}{\bar{\Gamma}  (t_{\cal H})} -1,
\end{equation}
and the new background parameters
\begin{equation}
\label{eq:delta_T_alpha}
\tilde{\Omega}_i = (1+w_i) \Omega_i,
\qquad
\tilde{\Omega} = \sum_i \tilde{\Omega}_i = 1+w_{\rm tot} \,.
\end{equation}

In the following, we consider a few particular cases which will be  relevant for our study of isocurvature fluctuations generated in the curvaton model.

\subsection{Case with $\Gamma =$ constant}
\label{sec:Gamma_const}

In  the standard curvaton scenario, $\Gamma$ is supposed to be constant. We can then easily recover the well-known results for the final density perturbation by setting $\delta_\Gamma =0$ in Eq.~\eqref{eq:deltaN_general}.  Before the curvaton decay, we have to take into account two fluids: the radiation originating from the inflaton and the oscillating curvaton,  which we treat  as  non-relativistic matter. We thus find
\begin{equation}
\delta N_{\cal H} = \frac{3 \Omega_\sigma \zeta_\sigma + 4 \Omega_r \zeta_r}{3 \Omega_\sigma + 4 \Omega_r},
\end{equation}
where $\Omega_i$ and $\zeta_i$ are  evaluated just before the decay. Since the total curvature perturbation $\zeta$ after the curvaton 
decay is given by $\zeta = \delta N_{\cal H}$, one obtains
\begin{equation}
\label{mixed_curvaton_inflaton}
\zeta= \frac{3 \Omega_\sigma \zeta_\sigma + 4 \Omega_r \zeta_r}{3 \Omega_\sigma + 4 \Omega_r}\,.
\end{equation}
In the original curvaton scenario, the perturbation of inflaton-generated radiation $\zeta_r$ is neglected, so that 
\begin{equation}
\label{eq:zeta_curvaton}
\zeta = r_{\rm dec} \, \zeta_\sigma, 
\end{equation}
where $r_{\rm dec}$ is defined as 
\begin{equation}
\label{eq:r_dec}
r_{\rm dec} = \left. \frac{3 \Omega_\sigma }{3 \Omega_\sigma + 4 \Omega_r} \right|_{\rm dec}\,.
\end{equation}
The more general formula (\ref{mixed_curvaton_inflaton}) applies to the mixed curvaton and inflaton scenarios \cite{Langlois:2008vk,Langlois:2004nn,Moroi:2005kz,Moroi:2005np,Ichikawa:2008iq,Enqvist:2013paa}. 

In order to compare the above analytic formula with  numerical calculations,  it is convenient to replace $r_{\rm dec}$  with the  quantity \cite{Kitajima:2014xna}:
\begin{equation}
\label{eq:r_s}
r_s \equiv \frac{3 \rho_{r\sigma, f} }{3 \rho_{r\sigma, f} + 4 \rho_{r\phi, f}},
\end{equation}
where $\rho_{r\sigma, f}$ and $\rho_{r\phi, f}$ are the energy densities of radiations generated by 
the curvaton and the inflaton, respectively, evaluated well after the curvaton decay (with the `$f$' subscript for ``final''). Indeed, 
when one numerically follows the evolution of radiation and of the curvaton,  one cannot define precisely  the time of the curvaton decay. Some possible choices for the time of the curvaton decay are when $H = \Gamma$, or $\Gamma t = 1$ where $t$ is the cosmic time, but  these two choices, which give different values for $r_{\rm dec}$, are somewhat arbitrary.  By contrast,  $r_s$ defined in Eq.~\eqref{eq:r_s} 
does not depend on the decay time and thus avoids any ambiguity.

For later convenience, we also note that $r_s$ is related to  the amount of  entropy produced during the curvaton decay, which can be expressed by the ratio
\begin{equation}
\label{def_Qs}
Q_s = \frac{\mathcal{S}_f}{\mathcal{S}_i},
\end{equation}
where $\mathcal{S}_f$ and $\mathcal{S}_i$ are entropies in a comoving volume respectively before and after the curvaton decay. 
Indeed, $r_s$ can be written as 
\begin{equation}
r_s = \frac{3(Q_s^{4/3}-1)}{ 3 Q_s^{4/3} + 1}.
\end{equation}

\subsection{Case with $\Gamma = \Gamma (T)$}
\label{sec:Gamma_of_T}

When we consider the freeze-out of DM particles such as WIMPs, the reaction rate is given by 
$\Gamma = \left\langle \sigma_{\rm int} v \right\rangle n_{\rm CDM}$ where $\sigma_{\rm int}$ is the interaction cross section,  $v$ the   relative velocity and  $n_{\rm CDM}$  the WIMP number density. In this case, $\Gamma$ depends on the temperature. For axionic dark matter, the transition occurs when the axion  field starts to oscillate, i.e. when the Hubble parameter drops below the axion mass, $m$, which in principle depends on the temperature. In this case, we have  $\Gamma = m(T)$.

Let us first derive  a general expression that applies  when $\Gamma$ depends on the temperature, before  
discussing separately the  particular cases of thermally decoupled DM and of axionic DM. 
Assuming that $\Gamma = \Gamma (T)$, one can write the fluctuations of $\Gamma$ as 
\begin{equation}
\label{eq:delta_Gamma_delta_T}
\delta_\Gamma = \alpha \, \delta_T,
\end{equation}
where 
\begin{equation}
\delta_T = \frac{\delta T}{T},
\qquad
\alpha = \frac{d \ln \Gamma}{d\ln T}.
\label{alpha}
\end{equation}
Note that the above parameter $\alpha$, whose explicit expression is model dependent,  has also been adopted in \cite{Lyth:2003ip}.
Since the temperature is related to  the radiation energy density as 
\begin{equation}
T \equiv \left[ \frac{30}{\pi^2 g_\ast } \rho_r \right]^{1/4},
\end{equation}
where $g_\ast$ is the effective degrees of freedom, the temperature can be written as 
\begin{equation}
\label{eq:T_pert}
T(t, \vec{x}) = \bar{T} (t) e^{\zeta_r - \delta N}\,,
\end{equation}
which implies 
\begin{equation}
\label{eq:delta_T}
\delta_T = \zeta_r - \delta N.
\end{equation}
Inserting the relation~\eqref{eq:delta_Gamma_delta_T} and \eqref{eq:T_pert} into Eq.~\eqref{eq:deltaN_general}, we arrive at
\begin{equation}
\label{eq:deltaN_H}
\delta N_{\cal H} = \frac{1}{ 3 \tilde{\Omega} - 2 \alpha} \left( 3 \sum_i \tilde{\Omega}_{i}\,  \zeta_{i} - 2 \alpha\,  \zeta_r \right).
\end{equation}

Let us now discuss the  fluctuations of DM. Regardless of whether DM particles are relativistic or non-relativistic, their number is conserved after the DM transition 
such as the freeze-out and 
the number density of DM particles $n_c$  satisfies the following equation of motion in each homogeneous patch:
\begin{equation}
\frac{dn_c}{dt} + 3 H n_c = 0\,.
\end{equation}
This implies that the local DM particle number density can be written as 
\begin{equation}
\label{eq:n_c}
n_c (t, \vec{x}) = \bar{n}_c (t) e^{3 (\zeta_c -\delta N)},
\end{equation}
where $\zeta_c$ expresses the perturbation associated with the number density.
Since in general  $n_c$ depends on the temperature only, 
one can relate the fluctuation of $n_c$ to those of the temperature as
\begin{equation}
\label{eq:delta_n}
\frac{\delta n_c}{\bar{n}_c} =  \frac{\partial \ln \bar{n}_c}{\partial \ln \bar{T}} \, \delta_T = \nu \delta_T,
\end{equation}
where we have introduced the parameter  
\begin{equation}
\label{nu}
\nu \equiv  \frac{\partial \ln \bar{n}_c}{\partial \ln \bar{T}}.
\end{equation}
By evaluating the number density fluctuation on the transition hypersurface, expanding Eq.~\eqref{eq:n_c} at linear order, and substituting Eqs.~\eqref{eq:delta_T} and \eqref{eq:delta_n}, we finally obtain 
\begin{equation}
\label{eq:zeta_c}
\zeta_c = \frac13 \left[ \nu \zeta_r + (3 -\nu) \delta N_{\cal H} \right].
\end{equation}

Assuming that only CDM and radiation are present after the transition, the  CDM isocurvature perturbation $S_c$ is usually defined by 
\begin{equation}
S_c \equiv 3 (\zeta_c - \zeta)\,,
\end{equation}
where $\zeta$ is the curvature  perturbation for radiation, which coincides here with the adiabatic perturbation. In the curvaton model, $\zeta$ is given by Eq.~\eqref{eq:zeta_curvaton}.
By deriving the expression $\zeta_c$ for some particular scenario of CDM generation, one can then predict the corresponding isocurvature fluctuations.
As mentioned in the introduction, the isocurvature fluctuations are now severely constrained by the CMB observations and the amplitude of isocurvature perturbation can thus provide a crucial test for a variety of models.
In the next section, we discuss  thermally decoupled DM and  axionic DM, both in the context of  the curvaton model.

\bigskip
\section{CDM Isocurvature fluctuations in the curvaton scenario}
\label{sec:iso_DM}

\subsection{Thermally decoupled CDM}

In this subsection, we study isocurvature fluctuations associated with  
thermally decoupled CDM in the curvaton model.
As discussed in many papers (see e.g., \cite{Moroi:2002rd,Lyth:2002my,Lyth:2003ip,Moroi:2008nn,Takahashi:2009cx,Lemoine:2006sc}),  
a large CDM isocurvature perturbation is generated if CDM is created (or, equivalently, if its number density freezes out) {\it before} the curvaton decay, whereas  no CDM isocurvature perturbation is generated when CDM is created {\it after} the curvaton decay. 
Here we revisit this issue in detail by using, on the one hand,  the sudden transition formalism developed in the previous section and, on the other hand,  
the numerical computation to follow the evolution of the homogeneous system\footnote{
The perturbations observed today were on super-Hubble scales at the time of the curvaton decay in the early Universe.  For  super-Hubble perturbations, it is sufficient to  follow the homogeneous evolution of the various densities with slightly different initial conditions and compare them later  in order to deduce the  amplitude of  the perturbations,  
in the spirit of the $\delta N$ formalism.
}.

The equations of motion we need to solve are:
\begin{eqnarray}
\label{eq:CDM_system_H1}
&& H^2 = 
\frac{1}{3 M_{\rm pl}^2} \left(\rho_r + \rho_\sigma + \rho_{\rm CDM} \right)  
\simeq
\frac{1}{3 M_{\rm pl}^2} \left(\rho_r + \rho_\sigma  \right) ,
\\
\label{eq:CDM_system_rho}
&& \frac{d\rho_r}{dt} + 4 H \rho_r = \Gamma \rho_\sigma,  
\\ 
\label{eq:CDM_system_rho}
&& \frac{d\rho_\sigma}{dt} + 3 H \rho_\sigma = - \Gamma \rho_\sigma,
\\ 
\label{eq:CDM_system_n_1}
&& \frac{d n_{\rm CDM}}{dt} + 3 H n_{\rm CDM} =  - \lambda \left( n_{\rm CDM}^2 - \left( n_{\rm CDM}^{\rm (eq)} \right)^2 \right),
\end{eqnarray}
where 
$\rho_r$, $\rho_\sigma$ and $\rho_{\rm CDM}$ are respectively the radiation,  curvaton and
 CDM energy densities, and
$\Gamma$ is the decay rate of the curvaton. 
The first equation above is the Friedmann equation, where $\rho_{\rm CDM}$ can be neglected since we are in the very early Universe. The second and third equations describe, respectively, the evolution of the radiation and curvaton energy densities, 
where we assume that the curvaton behaves as  non-relativistic matter. Their right-hand side characterizes the energy transfer between the curvaton and radiation as the curvaton decays. 

The last equation,  Eq.~\eqref{eq:CDM_system_n_1}, describes the evolution of the CDM particle number density $n_{\rm CDM}$, where, on the right hand side, $\lambda$ denotes the thermally-averaged annihilation cross section of CDM particles: $\lambda = \left\langle \sigma_{\rm ann} v \right\rangle$ with 
$\sigma_{\rm ann}$ and $v$ being the cross section and relative velocity, respectively.
The annihilation cross section can depend on the temperature 
as $ \left\langle \sigma_{\rm ann} v \right\rangle  \propto T^{q/2}$ where the cases with $q=0$ and $2$ correspond to $s$-wave and $p$-wave 
annihilations, respectively. 
The right hand side also depends on the number density at equilibrium, 
$n_{\rm CDM}^{\rm (eq)}$,  which is given, for non-relativistic particles,  by 
\begin{equation}
\label{eq:n_CDM_eq}
n_{\rm CDM}^{\rm (eq)} = g_{\rm CDM} \left( \frac{mT}{2\pi} \right)^{3/2} e^{-m/T},
\end{equation}
where $m$ is the mass of a CDM particle and $g_{\rm CDM}$ is the number of internal degrees of freedom.

\subsubsection{Instantaneous transition formalism}
\label{sec:ITF}

When we consider thermally decoupled DM such as WIMPs, the freeze-out phase when 
DM particles decouple from the thermal bath corresponds to the transition  after which 
the number of CDM particles is conserved. 
We define the instantaneous transition hypersurface  by the relation
\begin{equation}
H =\Gamma= \left\langle \sigma_{\rm ann} v \right\rangle n_{\rm CDM} \,,
\end{equation}
where the number density at freeze-out can be approximated 
 by the equilibrium expression  \eqref{eq:n_CDM_eq}. 
Taking into account  the possible temperature dependence of the annihilation cross section, 
$ \left\langle \sigma_{\rm ann} v \right\rangle  \propto T^{q/2}$, the parameter $\alpha$ defined in  \eqref{alpha}  is thus  given by 
\begin{equation}
\label{eq:alpha_wimp}
\alpha = \frac{m}{T} + \frac32 + \frac{q}{2}\,.
\end{equation}
Note that  $m/T \sim 20$ at  freeze out in the standard WIMP case.
The parameter  $\nu$ defined in Eq.~(\ref{nu}) is given by 
\begin{equation}
\label{eq:nu_wimp}
\nu = \frac{m}{T} + \frac32.
\end{equation}
Using this relation and the expression for $\delta N_{\cal H}$ in Eq.~\eqref{eq:deltaN_general}, one obtains 
\begin{equation}
\zeta_c =  \frac13 \left. \left[ 
\left( \frac{m}{T} + \frac32 \right)  \zeta_{r} 
+ \frac{1}{3 \tilde{\Omega} - 2 \alpha} \left( \frac32 - \frac{m}{T} \right)  \left( 3 \sum_i \tilde{\Omega}_{i} \zeta_{i} - 2 \alpha \zeta_r \right)
 \right] \right|_{\fr},
\end{equation}
where all  quantities on the right hand side are evaluated at the time of  freeze-out, as indicated by the subscript `fr'.

In the scenarios where only radiation and the curvaton are present, the above expression  reduces to 
\begin{equation}
\label{eq:zeta_c_wimp_curvaton}
\zeta_c =  \frac13 \left.  \left[ 
\left( \frac{m}{T} + \frac32 \right)  \zeta_r + \frac{1}{4 - 2 \alpha - \Omega_\sigma} \left( \frac32 - \frac{m}{T} \right)  
\left\{  3 \Omega_\sigma \zeta_\sigma + \left( 4 - 4 \Omega_\sigma - 2 \alpha \right) \zeta_r \right\}
 \right] \right|_{\fr}.
\end{equation}
In  the simplest curvaton models, one ignores the fluctuations of radiation produced by  the inflaton. Substituting $\zeta_r=0$ in \eqref{eq:zeta_c_wimp_curvaton}, one thus gets
\begin{equation}
\label{eq:S_c_curvaton}
S_c / \zeta = 3
\left[  
 \frac{\Omega_{\sigma, \fr}}{r_{\rm dec}} 
 \left( \frac32 - \frac{m}{T_\fr} \right)
\frac{1}{4 - 2 \alpha_\fr - \Omega_{\sigma, \fr}}   -1
\right],
\end{equation}
where the subscript `\fr' means that the corresponding quantity is evaluated at the time of freeze-out. 
This equation has already been derived in \cite{Lyth:2003ip}.

Eq.~\eqref{eq:S_c_curvaton}   works rather well when the
radiation produced by the curvaton decay is negligible at the time of  freeze-out. 
However, in comparison with the radiation produced by the inflaton decay,  the radiation generated by the curvaton decay can become significant.
When the freeze-out occurs after the curvaton dominates the Universe, the latter type of radiation  
constitutes most of the radiation component. In this case,  neglecting the radiation perturbation is not valid anymore since the perturbation of the curvaton-generated radiation  is similar to the initial curvaton perturbation. We discuss this point in more detail  in  the numerical study that follows.  In  Appendix~\ref{sec:dillute_plasma}, we also present a  modified version of our formalism where we artificially separate the two types of radiation and treat them as distinct components  ($\rho_r = \rho_{r\phi} + \rho_{r\sigma}$ where 
$\rho_r$ is the total radiation energy density and 
$\rho_{r\phi}$ and $\rho_{r\sigma}$ are respectively  the inflaton- and curvaton-generated radiation),
 which can give some insight about how the CDM perturbation and  the isocurvature fluctuations evolve.

Before moving to the numerical study, let us briefly mention  two limiting cases,  which have already been discussed in the literature:
\bigskip

\noindent
$\bullet$ {\bf Case where the freeze-out occurs  after the curvaton decay} \vspace{1mm} \\
In this case, by putting $\Omega_\sigma =0$ in  Eq.~\eqref{eq:zeta_c_wimp_curvaton}, we obtain $\zeta_c =\zeta_r$. Therefore, 
when the freeze-out occurs  after the curvaton decay, one finds
\begin{equation}
{S_c}= 0.
\end{equation}
No isocurvature fluctuations are generated in this case.

\bigskip
\noindent
$\bullet$ {\bf Case where the freeze-out occurs during RD epoch (the curvaton never dominates the Universe before the freeze-out)} \vspace{1mm} \\
When the freeze-out occurs during RD epoch and the curvaton never dominates the Universe before the freeze-out, 
we can put $\Omega_\sigma=0$. In addition, since radiation is essentially the product of the inflaton decay in this case, we can assume  $\zeta_r =0$ (neglecting the inflaton fluctuations) and Eq.~\eqref{eq:zeta_c_wimp_curvaton} yields $\zeta_c = 0$. Therefore, we obtain
\begin{equation}
\frac{S_c}{\zeta} = -3,
\end{equation}
which is completely excluded by current observations.

\subsubsection{Numerical evolution}

We now investigate the isocurvature fluctuations more precisely by resorting to a numerical calculation based on the $\delta N$ formalism. 
 In this calculation, we numerically solve
the background equations along  two trajectories with  different initial values for the curvaton energy density.  After solving the background equations, we evaluate the difference of the energy density of each component (radiation and DM) at a final time when the curvaton has completely decayed into  radiation. 
This final time is defined  on a hypersurface with a uniform $e$-folding number  (see Eq.~\eqref{eq:zeta_i}).

\begin{figure}[htbp]
\begin{center}
\includegraphics[width=150mm]{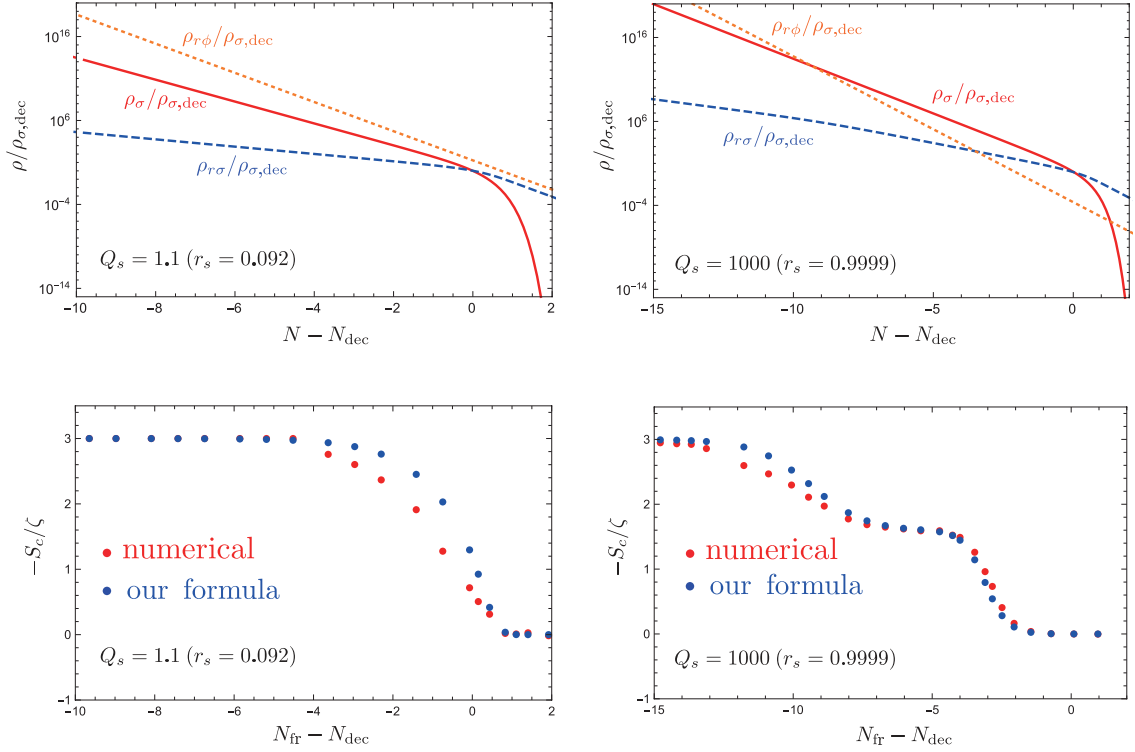}
\end{center}
\caption{The upper panels show the evolution of background energy density of 
the curvaton $\rho_\sigma$ (red solid), the inflaton-generated radiation $\rho_{r\phi}$ (orange dotted), and 
the curvaton-generated radiation $\rho_{r\sigma}$ (blue dashed),
as a function of $e$-folding number measured from the decay time of the curvaton ($N= N_\dec$).
In this plot,  $N_\dec$ corresponds to the time $t_\dec=\Gamma^{-1}$.
The lower panels show 
the residual CDM isocurvature perturbations as a function of the time of CDM freeze out ($N = N_
\fr$) measured from $N_\dec$. The red points in these panels are numerical results 
and  the blue points are  the ones calculated semi-analytically from Eq.~(\ref{eq:zeta_c_wimp_curvaton}). Left panels are for the case with $Q_s = 1.1$ and right panels
are for the case with $Q_s = 1000$. In this calculation, we fix $\lambda$ and the freeze-out temperature is changed by varying $m$. The temperature dependence of the annihilation cross section is fixed to be $q = 0$.}
\label{fig:isocurv}
\end{figure}

The background evolution of the energy densities $\rho_\sigma$ (red solid), $\rho_{r\phi}$ (orange dotted), and $\rho_{r\sigma}$ (blue dashed), in terms  of the number of $e$-folds, is plotted in the upper panels of Fig.~\ref{fig:isocurv} for two cases: in the left panel ($Q_s=1.1$), the curvaton always remains subdominant, whereas in the right panel ($Q_s = 1000$) it dominates the energy budget before its decay.

\begin{figure}[htbp]
\begin{center}
\includegraphics[width=150mm]{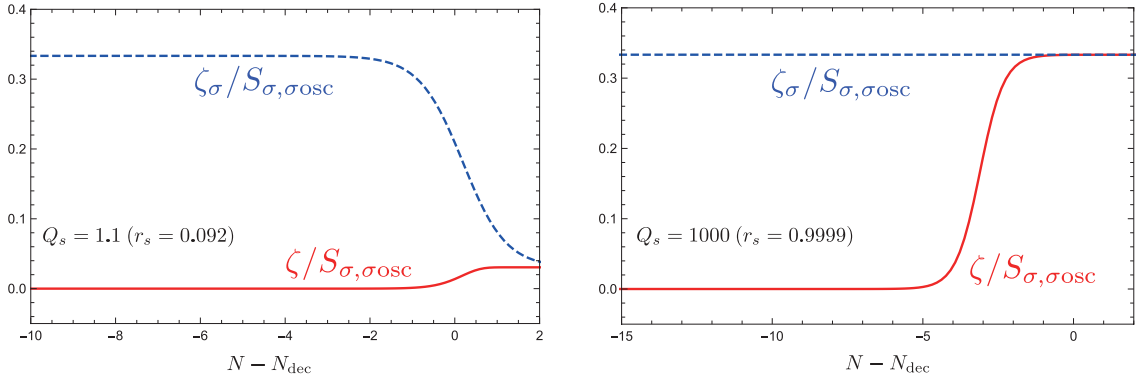}
\end{center}
\caption{The evolution of $\zeta$  (red line) and $\zeta_\sigma$ (blue dashed line) for $Q_s = 1.1$ (left panel) and $Q_s = 1000$ (right panel), as functions of $N - N_\dec$. Here the vertical axis is normalized by $S_{\sigma, \sigma\osc}$ which is the isocurvature perturbation of the curvaton ($S_\sigma := 3 (\zeta_\sigma - \zeta)$) at the time of
the onset of the curvaton oscillations (defined by $m_\sigma = H(N_{\sigma \osc})$).}
\label{fig:zeta_rsigma}
\end{figure}

In the  lower panels of Fig.~\ref{fig:isocurv}, we plot the residual CDM isocurvature perturbation,  or more precisely the quantity  $-{S_c}/{\zeta}$, as a function of the number of $e$-folds between  the CDM freeze-out  and the curvaton decay, i.e. $N_\dec-N_\fr$ (where the curvaton decay time is defined by $t_\dec=\Gamma^{-1}$ and the freeze-out time by $H=\lambda n_{\rm CDM}^{\rm (eq)}$).
The red points correspond to the fully numerical results, while the blue points are obtained with the analytical prediction from  Eq.~(\ref{eq:zeta_c_wimp_curvaton}). 
The blue points should be regarded as resulting from a semi-analytic calculation, since we use  the formula (\ref{eq:zeta_c_wimp_curvaton}) with the values  of $\zeta_r(N_\fr)$ and $\zeta_\sigma (N_\fr)$ at freeze-out time, obtained from the numerical  evolution of $\zeta_r$ (red line) and $\zeta_\sigma$ (blue dashed line) as shown in  Fig. \ref{fig:zeta_rsigma}.
%

The lower panels of  Fig.~\ref{fig:isocurv} show that our formula is mostly consistent with the numerical result
in both cases.
 One can  clearly see the two limiting behaviours already discussed,  $S_c/\zeta\simeq -3$ and $S_c/\zeta\simeq 0$ when the freeze-out takes place, respectively, long before or long after the curvaton decay. 
In the case $Q_s = 1.1$, these two limits are separated by a transition of roughly $5$ $e$-folds. By contrast, 
in the case with $Q_s = 1000$, where the curvaton reaches domination  before its decay,
one can see that there exists an intermediate region where $-S_c / \zeta \approx 1.5$. Moreover, one notes that 
$-S_c / \zeta$ is strongly suppressed even for negative values of $N_\fr-N_\dec$, i.e.  even  when the freeze-out occurs before the completion of the curvaton decay.

By comparing the behavior of the residual CDM isocurvature perturbations with the background dynamics shown in the upper panel,
we can see that the range of $N_\fr - N_\dec$ where $-S_c / \zeta \approx 1.5$ corresponds to a phase when  the curvaton dominates the Universe while
 the radiation component is dominated by the inflaton-generated component.
If the freeze-out occurs during such an era,  one can substitute $\zeta_r = 0$ and $\Omega_{\sigma, \fr} \approx r_s \approx 1$ in the  expression \eqref{eq:zeta_c_wimp_curvaton}, which  implies~\cite{Lyth:2003ip}
\begin{eqnarray}
 - S_c / \zeta \approx - 3 \left[ \left( {3 \over 2} - {m \over T_\fr} \right) {1 \over 3 - 2 \alpha_\fr} - 1 \right] .
 \end{eqnarray}
Assuming $m/T_\fr   \gg 1$, and hence $\alpha_\fr  \approx m/T_\fr$ according to (\ref{eq:alpha_wimp}), yields
$S_c / \zeta \approx -3/2$.

Next let us focus on the region where $S_c / \zeta$ goes to zero  when the freeze-out occurs (slightly) before the curvaton decay.
This region corresponds to the era where the matter-like curvaton dominates the Universe while the radiation component is also dominated by the curvaton-generated one.
In this regime, the Universe is completely dominated by the two components associated with the curvaton and, in this sense, 
 evolves adiabatically. Therefore,  the residual CDM isocurvature perturbations also vanish in this limit.
 
Once the value of $Q_s$  is given, we can estimate the duration of the period during which  the radiation component is completely dominated 
by the curvaton-generated  radiation, before the curvaton decay. This duration can be represented by the quantity $N_\dec - N_{\rm dom}$  
as a function of $Q_s$,  where $N_{\rm dom}$ is defined as the time when the curvaton-generated radiation  starts to dominate the inflaton-generated radiation  
(i.e., $\rho_{r\sigma}(N_{\rm dom})=\rho_{r\phi}(N_{\rm dom})$)\footnote{
In the following, the subscript `dom' indicates that the quantity is evaluated at time of $N=N_{\rm dom}$.
}
When $Q_s$ is large, it  can be approximated as  
\begin{equation}
\label{eq:Q_s_large}
Q_s \simeq \left( \frac{\rho_{r \sigma}}{\rho_{r\phi}} \right)^{3/4},
\end{equation}
where the right-hand side is evaluated at the time of the curvaton decay. Since $\rho_{r \sigma}$ and $\rho_{r\phi}$ respectively scale as 
$\rho_{r \sigma} \propto a^{-3/2}$ (for the curvaton-dominated Universe) and $\rho_{r\phi} \propto a^{-4}$,
and using the definition of $N_{\rm dom}$, one can derive 
\begin{equation}
N_{\rm dom}  - N_{\rm dec} \simeq \frac{8}{15} \ln Q_s,
\end{equation}
which gives the time when the isocurvature fluctuations vanish once $Q_s$ is given.

\begin{figure}[htbp]
\begin{center}
\includegraphics[width=150mm]{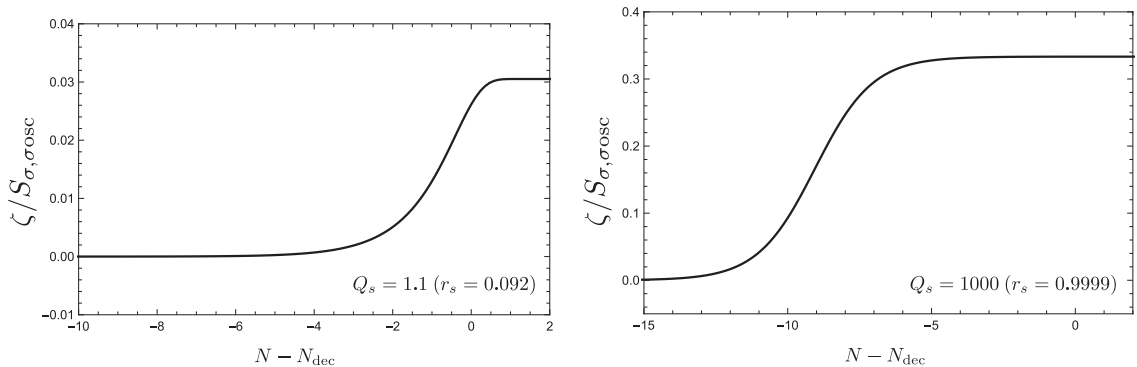}
\end{center}
\caption{The evolution of $\zeta$ as a function of $N-N_\dec$ for $Q_s = 1.1$ (left panel) and $Q_s = 1000$ (right panel). As in Fig.~\ref{fig:zeta_rsigma},
the vertical axis is normalized by $S_{\sigma,\sigma \osc}$.}
\label{fig:diff}
\end{figure}

\subsubsection{Difference between the analytical and numerical results}
Let us briefly discuss the  deviation between our analytic formula and the numerical result shown in Fig.~\ref{fig:isocurv}.
The sudden transition formalism (\ref{eq:zeta_c_wimp_curvaton}) uses the value of each curvature perturbation at  freeze-out 
but does not  depend on the information about how the perturbations evolve.
In order to show  that the time variation of $\zeta$ could be relevant to explain the deviation between the analytical and numerical results, 
we plot the evolution of the curvature perturbation $\zeta$, as a function of the number of $e$-folds $N - N_\dec$ in Fig.~\ref{fig:diff}.
Comparing this figure with Fig.~\ref{fig:isocurv}, one notices that $\zeta$ starts to evolve significantly in the regions where the deviation 
between the numerical result and the sudden transition formalism becomes large  as shown in Fig. \ref{fig:isocurv}.
Nevertheless  we should stress that the basic behavior of the CDM isocurvature perturbations can be understood by the use of the sudden transition formalism  (\ref{eq:zeta_c_wimp_curvaton}).

\subsection{Axion DM}

The role of DM can also be played by an oscillating scalar field, such as an axion \cite{Peccei:1977hh,Peccei:1977ur,Weinberg:1977ma,Wilczek:1977pj} (see also e.g. \cite{Kawasaki:2013ae,Marsh:2015xka} for recent reviews on cosmological aspects of axion DM). 
Denoting the axion field as $\chi$ and assuming that its potential is of the form $V (\chi) =m^2 \chi^2/2$ (at least at the bottom of the potential), 
 its equation of motion is given by
\begin{equation}
\ddot{\chi} + 3 H \dot{\chi} + m^2(T) \chi = 0,
\end{equation}
where $m(T)$ is the axion mass and can depend on the temperature above a critical temperature $T_{\rm cr} $: 
\begin{equation}
\label{eq:m_chi}
m(T) \simeq \begin{cases} m_\ast (T_{\rm cr}/T )^\beta ~~ &\text{for}~ T > T_{\rm cr} \\
m_\ast ~~ &\text{for}~ T < T_{\rm cr} \end{cases}.
\end{equation}
For the ordinary QCD axion, one has $\beta = 3.34$, $T_{\rm cr} = 0.26 \Lambda_{\rm QCD}$ with $\Lambda_{\rm QCD} = 400$ MeV   and $m_* = 3.82 \times 10^{-2} \Lambda_{\rm QCD}^2/F_a\simeq 6 \times 10^{-6}~{\rm eV} (10^{12}~{\rm GeV}/F_a)$,  $F_a$ being the axion decay constant \cite{Wantz:2009it}.

The production of axions occurs via the so-called misalignment mechanism. The initial amplitude of the axion oscillations is given by $\chi_i = F_a \theta_i$, where $\theta_i$ is an initial misalignment angle.
Just after the axion field begins to oscillate, the axion number density is given by 
\begin{equation}
\label{n_chi}
n_\chi =  \frac{1}{2m(T)} (\dot\chi^2 + m^2 (T)  \chi^2),
\end{equation}
and then evolves adiabatically as $n_\chi \propto a^{-3}$.  

One can consider that axion DM is ``created" at the sudden transition hypersurface that corresponds to the onset of  the axion oscillations, defined by 
\begin{equation}
\label{eq:osc_begin}
 H (N_{\chi\osc}) = m (N_{\chi\osc})\,,
\end{equation}
where $N_{\chi\osc}$ denotes the number of $e$-folds at the onset of the  axion oscillations.
From the above temperature-dependence of the axion mass, given in Eq.~(\ref{eq:m_chi}), we can easily calculate the coefficients $\alpha$ and $\nu$, 
introduced in Eqs.~(\ref{alpha}) and (\ref{nu}) respectively, and we get
\begin{equation}
\label{eq:alpha_nu_axion}
\alpha = \nu = -\beta.
\end{equation}
Substituting these relations into Eqs.~(\ref{eq:deltaN_H}) and (\ref{eq:zeta_c}),
   the DM perturbation is found to be given by 
\begin{equation}
\zeta_c = \frac13 \left. \left[
-\beta \zeta_r + \frac{3+\beta}{3 \tilde{\Omega} + 2 \beta} \left( 3 \sum_i \tilde{\Omega}_{i} \zeta_{i} + 2 \beta \zeta_r \right)
\right] \right|_{\chi\osc}\,,
\label{eq:zeta_c_axion}
\end{equation}
where the quantities on the right hand side are evaluated at the onset of the axion oscillations.

In the context of the curvaton scenario, where only radiation and the curvaton fluid are present, 
the above formula  reduces to
\begin{equation}
\zeta_c = \frac13 \left. \left[
-\beta \zeta_r + \frac{3+\beta}{4 + 2\beta -  \tilde{\Omega}_\sigma} 
\left\{  3 \Omega_\sigma \zeta_\sigma + \left( 4 - 4 \Omega_\sigma + 2 \beta \right) \zeta_r \right\} 
\right] \right|_{\chi\osc}.
\end{equation}
By putting $\zeta_r=0$, which is assumed in the original curvaton model, one can write the isocurvature fluctuation for the axion DM as 
\begin{equation}
\label{eq:S_c_axion}
S_c /\zeta =
 3 \left(  \frac{\Omega_{\sigma, \chi\osc}}{r_\dec} \frac{ 3 +  \beta }{4 + 2 \beta - \Omega_{\sigma, \chi\osc}}  -1 \right).
\end{equation}
This formula has also been obtained in \cite{Lyth:2003ip}. However, as we already stressed in the WIMP case, 
the assumption of $\zeta_r=0$ is not really justified when radiation is dominated by its curvaton generated component  at the onset of the axion oscillations. 

\begin{figure}[htp]
\centering
\subfigure{
\includegraphics [width = 7.5cm]{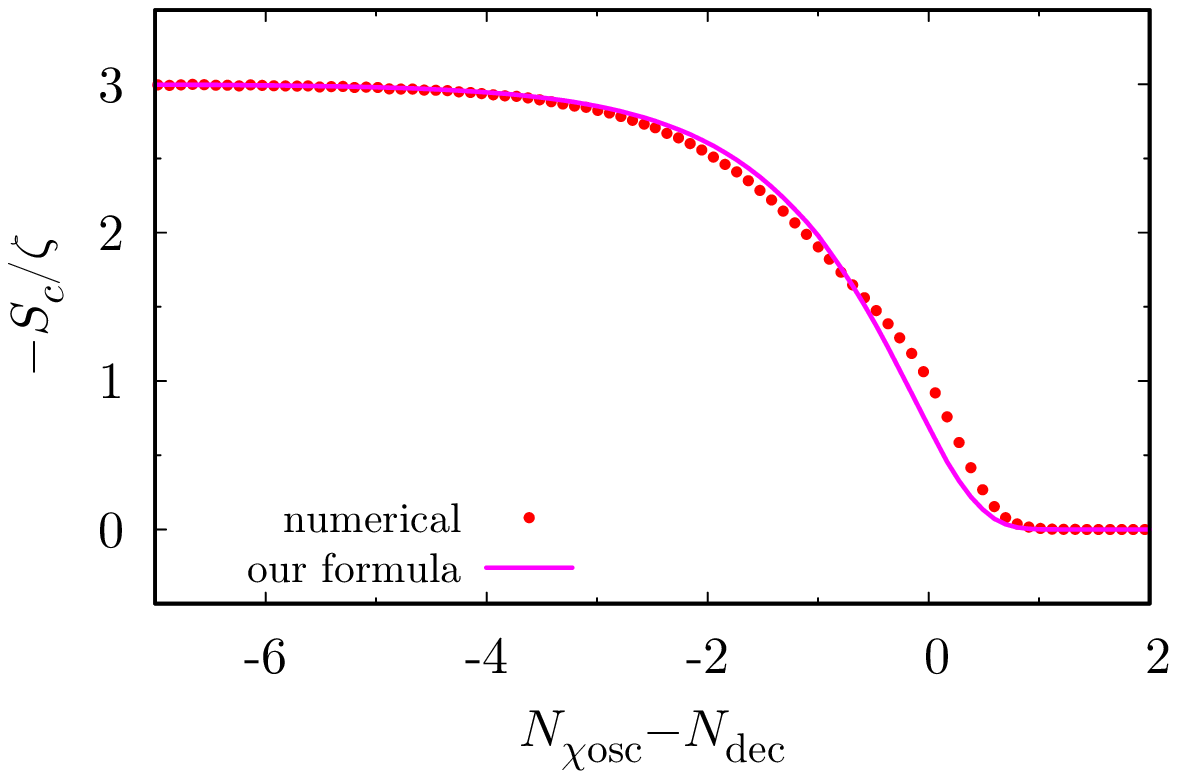}
\label{subfig:axion_iso_1}
}
\subfigure{
\includegraphics [width = 7.5cm]{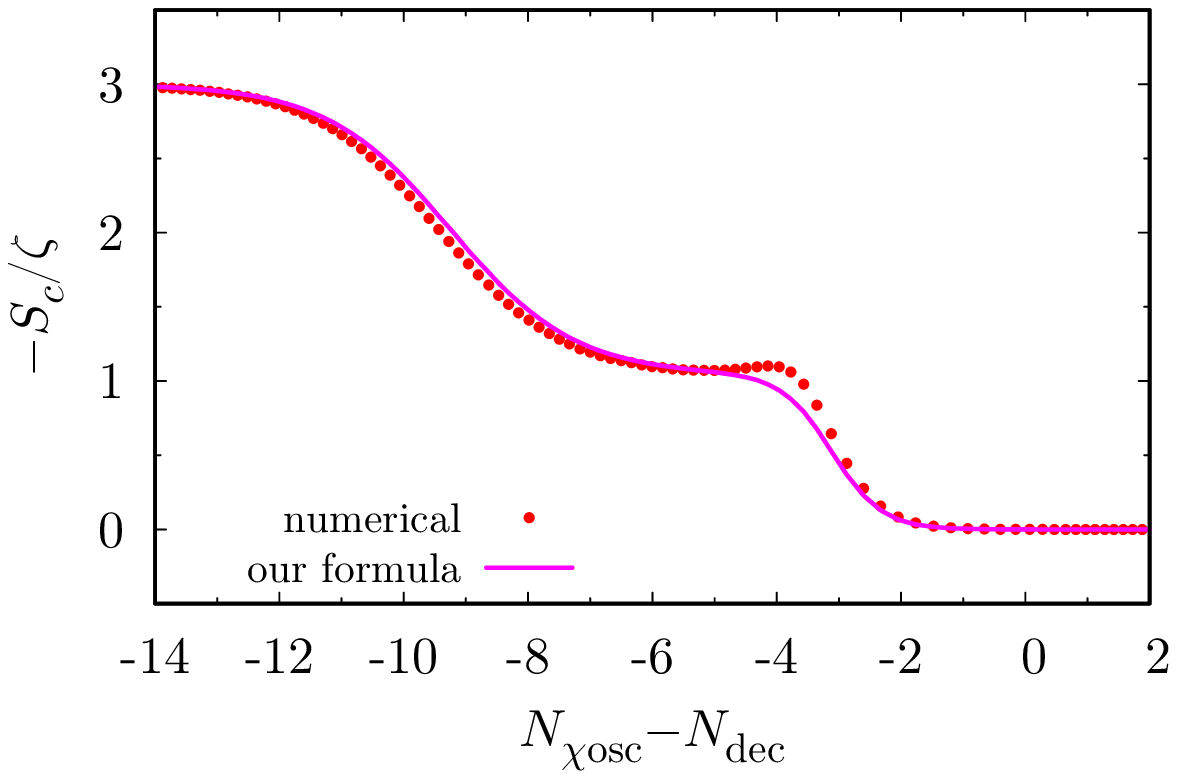}
\label{subfig:axion_iso_2}
}
\vspace{1cm}
\caption{
Plot of $-S_c/\zeta$ as a function of $N_{\chi\osc} - N_\dec$ for the cases with $\beta = 3.34$ and $F_a = 10^{12}$~GeV and $m_\sigma = 10^6$ GeV.
The magenta line represents the semi-analytic formula (\ref{eq:zeta_c_axion}).
We show the cases with $Q_s = 1.1$ (left), and 1000 (right). 
Here $N_{\chi\osc}$ and $N_{\rm dec}$ are the $e$-folding numbers at $H=m(T)$ and $\Gamma t = 1$ respectively. 
}
\label{fig:axion_iso}
\end{figure}

\begin{figure}[htbp]
\begin{center}
\includegraphics[width=8.5cm]{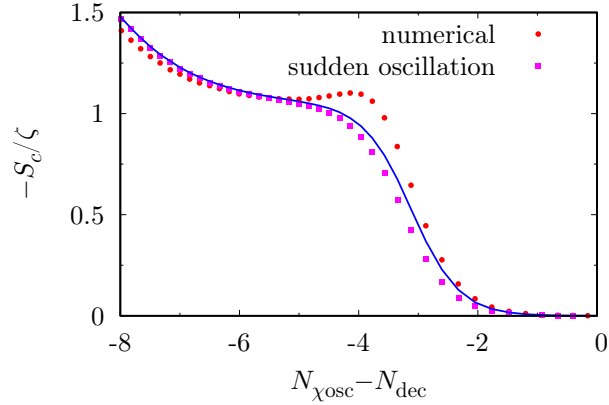}
\end{center}
\caption{
Comparison between the numerical result in Fig.~\ref{subfig:axion_iso_2} (red circles) and the ``sudden oscillation" 
case where the axion number density is set to  evolve as $n_\chi \propto a^{-3}$
just after the epoch of $H = m(T)$  (magenta squares).}
\label{fig:sudden_osc}
\end{figure}

Fig.~\ref{fig:axion_iso} shows the comparison between our numerical results (red points) and the  semi-analytic formula (\ref{eq:zeta_c_axion}) using the values of $\zeta_r$ and $\zeta_\sigma$ obtained numerically, with the following choice of parameters: $\beta = 3.34$, $F_a = 10^{12}$ GeV and $m_\sigma = 10^6$ GeV with $Q_s = 1.1$ (left), and 1000 (right).
While the numerical result presents an overall good agreement with our analytic formula, there is nevertheless some difference between them. In particular, a bump appears near $\Delta N\equiv N_{\chi \osc}-N_{\rm dec} \simeq 4$ in the case  of $Q_s = 1000$ (right panel of Fig.~\ref{fig:axion_iso}). 
This is due to the fact that DM is an  oscillating scalar field. 
In the sudden transition approximation, we assume that the axion comoving number density is instantaneously fixed at the beginning of the axion oscillations and then conserved. However, this conservation is strictly valid only for an adiabatic evolution, which takes place later when $m\gg H$. In practice, this condition is not yet fully satisfied just after the onset of oscillations.
Thus, the final value is sensitive to the change of the oscillation frequency (or time derivative of the axion mass, $\dot{m}$) at that time.
It becomes maximal when the equation of state for the radiation component changes significantly due to the curvaton generated radiation. 
Fig.~\ref{fig:sudden_osc} shows the difference between the realistic case (the numerical result in Fig.~\ref{subfig:axion_iso_2}) and an idealized case where the axion number density is initially determined by the expression (\ref{n_chi}),  at the time when $H = m(T)$,  and then assumed to  evolve  as $n_\chi \propto a^{-3}$. 
The latter case clearly satisfies the adiabatic condition and, as a consequence, the bump does not appear.

\section{Implications for the constraints on model parameters}
\label{sec:implications}

We now discuss the implications of our results concerning the constraints on the model parameters in the curvaton scenario.
So far, it has been argued in the literature  that, when the transition that produces DM  
occurs before the curvaton  decay, sizable isocurvature fluctuations are generated, which is inconsistent with observations. 
However, as shown in the previous section, isocurvature fluctuations can be suppressed even
 when the freeze-out occurs before the curvaton decay 
 provided the radiation generated by the curvaton decay  is the main component  in the radiation.  

In this section, we discuss the constraints on the initial amplitude of the curvaton $\sigma_\ast$ and  its decay rate $\Gamma$ that can be deduced from the non-detection of isocurvature fluctuations. Indeed,  the isocurvature fluctuations are severely constrained by the present CMB observations, such as those of Planck. 
Isocurvature and adiabatic perturbations produced in multi-field inflationary scenarios can be correlated, as shown in \cite{Langlois:1999dw}, and the CMB  constraints depend on the degree of correlation between the isocurvature and the adiabatic perturbations.   In the curvaton scenario  considered here, isocurvature perturbations are negatively correlated with adiabatic ones\footnote{
Generally in the curvaton scenario, 
where DM is produced from the curvaton decay, the so-called positively correlated isocurvature fluctuations are generated. By contrast,  when DM is not produced directly from the curvaton decay but from another sector, one gets negatively correlated isocurvature perturbations, as  the scenarios discussed in this paper.
}.
As a conservative estimate, we adopt here  the isocurvature upper bound  
\begin{equation}
\left| \frac{S_c}{\zeta} \right| < 0.1,
\end{equation}
from which we derive constraints on $\sigma_\ast$ and $\Gamma$.
In the following, we present our results for the  WIMP and axion cases, successively.

Before discussing the isocurvature constraints, it should be noted that these parameters can also be constrained by non-Gaussianity\footnote{
We consider here the non-Gaussianity of the dominant adiabatic mode. Note that the isocurvature modes  also lead 
to non-Gaussianities, which differ from the adiabatic ones, although they are usually subdominant except in special models 
\cite{Langlois:2011zz,Langlois:2010fe,Langlois:2011hn,Langlois:2012tm,Hikage:2012tf}.
}. 
Indeed, Planck observations have tightly constrained the so-called non-linearity parameter $f_{\rm NL}$.  
In the curvaton scenario, the non-Gaussianity is of  the local type and the corresponding limits are given by 
$f_{\rm NL}^{\rm (local)} = 2.5 \pm 5.7 $ (68 \% C.L.) \cite{Ade:2015ava}, from which one derives a lower bound on $r_{\rm dec}$:
\begin{equation}
\label{eq:rdec_const}
r_{\rm dec} > 0.16 
\qquad (68~\% {\rm C.L.}).
\end{equation}
Since this quantity can be approximated by  \cite{Lyth:2001nq}
\begin{equation}
r_{\rm dec} \sim {\rm Min} \left[1, \left( \frac{\sigma_\ast}{M_{\rm pl}} \right)^2  \left( \frac{m_\sigma}{\Gamma} \right)^{1/2} \right] \,,
\end{equation}
 where $\sigma_\ast$ is the initial amplitude of the curvaton,
we get that  the allowed region in the $\sigma_\ast$--$\Gamma$ is characterized by
\begin{equation}
{\Gamma \over M_{\rm pl}}  \lesssim 10^{-13}\left( {m_\sigma \over 10^6~ {\rm GeV}} \right) \left( {\sigma_* \over M_{\rm pl}}\right)^4~.
\end{equation}
We will take into  account the constraint from non-Gaussianity  when we show our results on the isocurvature constraints.

\subsection{WIMP case}
\begin{figure}[htbp]
\begin{center}
\includegraphics[width=100mm]{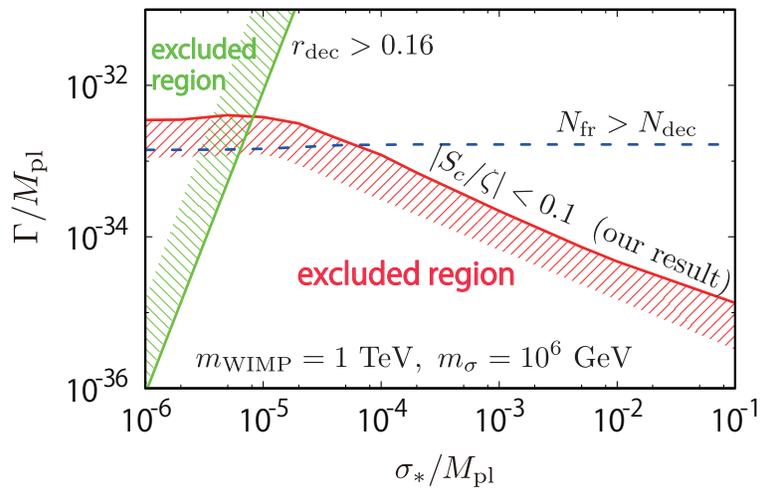}
\end{center}
\caption{Excluded region from the isocurvature constraints along the one from non-Gaussianity (green region) in the $\sigma_*$ - $\Gamma$ space. The red region is our result 
and the blue one  corresponds to the conventional constraint $N_{\rm fr} > N_{\rm dec}$. Here, we have chosen the illustrative values $m_{\rm WIMP} = 1$~TeV,
$m_{\sigma} = 10^6$~GeV, and  have tuned the thermally-averaged annihilation cross section of WIMP so that $\Omega_{\rm WIMP} h^2 = 0.12$. The allowed region
is top right one.
}
\label{fig:iso_const}
\end{figure}

We start with  the case of WIMP dark matter.
In Fig.~\ref{fig:iso_const},  we show the region excluded  by the isocurvature constraint, together 
 with the one from non-Gaussianity (green region) given by Eq.~\eqref{eq:rdec_const}.
Here, we have taken $m_{\rm WIMP} = 1$~TeV,
$m_{\sigma} = 10^6$~GeV, and  have tuned the thermally-averaged annihilation cross section of WIMP, $\lambda$, so that $\Omega_{\rm WIMP} h^2 = 0.12$~\cite{Ade:2015xua}.

When the curvaton dominates the total energy density before its decay,
 the final abundance of DM is  modified from the one calculated in the standard scenario.
 Here we derive an approximate  expression for the theoretically-expected $\Omega_{\rm WIMP}$ in such a case.
Freeze-out is characterized by 
\begin{eqnarray}
H_\fr = \lambda \, n^{\rm (eq)}_{{\rm WIMP}, \fr}\,,
\label{eq:freeze}
\end{eqnarray}
and
\begin{eqnarray}
H_\fr^2 = {\rho_{r,\fr} \over 3 M_{\rm pl}^2} \left( 1 - \Omega_{\sigma, \fr} \right)^{-1}\,.
\label{eq:hubfr}
\end{eqnarray}

Taking account of the entropy production from the curvaton, in the case where the freeze-out occurs before the curvaton decay, that is, $T_\fr > T_\dec$,
the density parameter of WIMP at the present time can be evaluated as
\begin{equation}
\Omega_{\rm WIMP} h^2 = \frac{m_{\rm WIMP} n_{\rm WIMP,0}}{\rho_{\rm crit}/h^2}
= m_{\rm WIMP} \frac{n_{\rm WIMP, 0}}{s_0} \frac{s_0}{\rho_{\rm crit}/h^2}
= m_{\rm WIMP} R_s^{-1} \frac{n_{\rm WIMP, fr}}{s (T_\fr)} \frac{s_0}{\rho_{\rm crit}/h^2},
\end{equation}
where $\rho_{\rm crit}$ is the critical energy density and $s_0$ is the entropy density at the present time.
$R_s$ is the dilution factor given by the ratio of the entropy at the freeze-out time and that  after the curvaton decay,
\begin{eqnarray}
R_s := {\mathcal{S}_f \over \mathcal{S}_\fr} = Q_s {\mathcal{S}_{i} \over \mathcal{S}_\fr },
\end{eqnarray}
where $Q_s$ is defined in Eq.~\eqref{def_Qs} and 
can be expressed as
\begin{equation}
\label{eq:Q_s}
Q_s = \frac{\mathcal{S}_f}{\mathcal{S}_i} 
= \max \left[1, \frac{4}{3T_{\rm dec}} \frac{\rho_\sigma}{s}\bigg|_{\osc} \right]
= \max \left[1,\frac16 \left( \frac{m_\sigma}{\Gamma_\sigma} \right)^{1/2}  \bigg(\frac{\sigma_*}{M_\pl}\bigg)^2 \right],
\end{equation}
where we have used 
\begin{equation}
\label{eq:Q_s_derive}
 \frac{4}{3T_{\rm dec}} \frac{\rho_\sigma}{s}\bigg|_{\osc} 
=\frac{T_\osc}{T_{\rm dec}} \Omega_{\sigma,\osc}  
=\frac{1}{6} \frac{T_\osc}{T_{\rm dec}} \bigg(\frac{\sigma_*}{M_\pl}\bigg)^2 
= \frac16 \left( \frac{m_\sigma}{\Gamma_\sigma} \right)^{1/2}  \bigg(\frac{\sigma_*}{M_\pl}\bigg)^2.
\end{equation}
By assuming the instantaneous decay of the curvaton at $t=t_\dec$,
that is, neglecting the contribution of the curvaton-generated radiation before the curvaton decay ($t = t_\dec$), we have
$\mathcal{S}_i = \mathcal{S}_\fr$ for the case with $T_\fr > T_\dec$, which implies  $R_s = Q_s$. 
However, once the curvaton-generated radiation is taken into account,
$\mathcal{S}_{i} / \mathcal{S}_\fr $ should depend on when the WIMP freeze-out occurs even before the curvaton decay.
By using  $\rho_{\rm crit}/s_0 h^2 = 3.64 \times 10^{-9}$ GeV and assuming that $g_{\ast s,  \fr}= g_{\ast, \fr} = 106.75$, we  get 
\begin{equation}
\label{eq:Omega_wimp_general}
\Omega_{\rm WIMP} h^2   
\simeq 0.12 
 \left( {m_{\rm WIMP} /T_\fr \over 20} \right) \left( {3.3 \times 10^{9}~ {\rm GeV}^{-1}  \over \lambda M_{\rm pl}} \right) 
\times
\begin{cases} 
1 ~~&\text{for}~~T_{\fr} < T_{\rm dec} \\ \left( 1 - \Omega_{\sigma, \fr}\right)^{-1/2}R_s^{-1} ~~&\text{for}~~T_{\fr} > T_{\rm dec} 
\end{cases}.
\end{equation} 
Here, we have used $n_{\rm WIMP,fr} = n_{\rm WIMP,fr}^{\rm (eq)} = H_\fr / \lambda$ based on Eqs.~(\ref{eq:freeze}) and (\ref{eq:hubfr}).
The expression for $T_\fr > T_{\rm dec}$ can be further divided into 3 cases, which we now discuss in turn.

\bigskip
\noindent
$\bullet$ {\bf Case 1 : freeze-out occurs during inflaton-generated radiation dominated era } 

When the WIMP freeze-out occurs during the inflaton-generated radiation domination  before the curvaton decay,
 we can take $\Omega_{\sigma,\fr} \to 0$ and $R_s = Q_s$ in Eq.~\eqref{eq:Omega_wimp_general}.
Then, $\Omega_{\rm WIMP}$ is simply given as
\begin{equation}
\Omega_{\rm WIMP} h^2   
\simeq 0.12 
 \left( {m_{\rm WIMP} /T_\fr \over 20} \right) \left( {3.3 \times 10^{9}~ {\rm GeV}^{-1}  \over \lambda M_{\rm pl}} \right) 
 \,Q_s^{-1}.
\end{equation}

\bigskip
\noindent
$\bullet$ {\bf Case 2 : freeze-out occurs during curvaton dominated era with $T_{\fr} > T_{\rm dom}$}

In the case where the freeze-out occurs during the curvaton domination,
we should take into account the factor  $(1 - \Omega_{\sigma,\fr})$.
Since here the freeze-out occurs before the curvaton-generated radiation becomes dominant in the total radiation component,  i.e., $T_\fr > T_{\rm dom}$, 
the factor $R_s$ should still be identical to $Q_s$, and hence  we have in  this  case
\begin{equation}
\label{eq:Omega2}
\Omega_{\rm WIMP} h^2   
\simeq 0.12 
 \left( {m_{\rm WIMP} /T_\fr \over 20} \right) \left( {3.3 \times 10^{9}~ {\rm GeV}^{-1}  \over \lambda M_{\rm pl}} \right) 
 \,\left( 1 - \Omega_{\sigma, \fr}\right)^{-1/2} Q_s^{-1},
\end{equation}  
where $(1 - \Omega_{\sigma,\fr})$ is given by 
\begin{eqnarray}
1 - \Omega_{\sigma,\fr} &=& {\rho_{r,\fr} \over 3 M_{\rm pl}^2 H_\fr^2} \simeq {\rho_{r,\fr} \over \rho_{\sigma,\fr}}
= {\rho_{r,{\rm osc}} \over \rho_{\sigma, {\rm osc}}} \left( {T_\fr \over T_{\rm osc}}\right) 
\cr\cr
&=& 6 \left({90 \over \pi^2 g_{\ast,{\rm osc}}} \right)^{-1/4}
\left( {\sigma_\ast \over M_{\rm pl}} \right)^{-2} \left( {m_\sigma \over M_{\rm pl}}\right)^{-1/2}\left( {m_{\rm WIMP} \over M_{\rm pl}}\right)
\left( {T_\fr \over m_{\rm WIMP}}\right).
\label{eq:Omegasig}
\end{eqnarray}

\bigskip
\noindent
$\bullet$ {\bf Case 3 : freeze-out occurs during curvaton dominated era with $T_{\fr} < T_{\rm dom}$}

When the freeze-out occurs after the domination of the  curvaton-generated radiation over the inflaton-generated radiation (but note that the total energy density of the Universe is still dominated by the curvaton), 
the factor $(1 - \Omega_{\sigma,\fr})$ becomes different from  Eq.~(\ref{eq:Omegasig})  and can be expressed as
\begin{eqnarray}
1 - \Omega_{\sigma,\fr} &=& {\rho_{r\sigma,\fr} \over 3 M_{\rm pl}^2 H_\fr^2} = {\rho_{r\sigma,\fr} \over \rho_{r\sigma,\fr} + \rho_{\sigma,\fr}}
\simeq { (T_\dec / T_\fr )^4 \over 1 + (T_\dec / T_\fr )^4 },
\label{eq:Omegasig2}
\end{eqnarray}
where we have used $\rho_{r \sigma, \dec} \simeq \rho_{\sigma, \dec}$, and the scaling  $\rho_r \simeq \rho_{r \sigma} \propto a^{-3/2}$ for $T_{\rm dom} > T_\fr >  T_\dec$. Moreover, we have
\begin{eqnarray}
{T_\dec \over T_\fr} \simeq \left( {90 \over \pi^2 g_{\ast,\dec}}\right)^{1/4}
\left( {m_{\rm WIMP} \over T_\fr}\right) \left( {m_{\rm WIMP} \over M_{\rm pl}} \right)^{-1}
\left( {\Gamma \over M_{\rm pl}} \right)^{1/2} .
\label{eq:Tfrdec}
\end{eqnarray}
Furthermore,  the factor $R_s$ is not identical to the total entropy production ratio $Q_s$ 
because  $\mathcal{S}_i / \mathcal{S}_\fr \neq 1$ due to the contribution of the curvaton-generated radiation.
By using the scaling for the energy density of the curvaton-generated radiation given by
$\rho_{r \sigma} \propto a^{-3/2}$~\cite{Kolb:text}, we find
\begin{eqnarray}
{\mathcal{S}_i \over \mathcal{S}_\fr} = {T_i^3 a_i^3 \over T_\fr^3 a_\fr^3} = {T_{\rm dom}^3 a_{\rm dom}^3 \over T_\fr^3 a_\fr^3}
= \left({a_{\rm dom} \over a_\fr} \right)^{-9/8} \left({a_{\rm dom} \over a_\fr} \right)^{3} = e^{-{15 \over 8} (N_\fr - N_{\rm dom})},
\end{eqnarray}
from which  we get  the relation 
\begin{eqnarray}
R_s =  e^{-{15 \over 8} (N_\fr - N_{\rm dom})}\, Q_s \,.
\end{eqnarray}
In conclusion,  we have for this case
\begin{equation}
\label{eq:Omega3}
\Omega_{\rm WIMP} h^2   
\simeq 0.12 
 \left( {m_{\rm WIMP} /T_\fr \over 20} \right) \left( {3.3 \times 10^{9}~ {\rm GeV}^{-1}  \over \lambda M_{\rm pl}} \right) 
 \,\left( 1 - \Omega_{\sigma, \fr}\right)^{-1/2} Q_s^{-1} \, e^{{15 \over 8} (N_\fr - N_{\rm dom})}.
\end{equation}  
The extra factor $e^{{15 \over 8} (N_\fr - N_{\rm dom})}$ can also be expressed in terms of  the model parameters as
\begin{eqnarray}
e^{{15 \over 8} (N_\fr - N_{\rm dom})} &=& \left( {a_\fr \over a_{\rm dom}}\right)^{15/8} = \left( {T_\fr \over T_{\rm dom}} \right)^{-5} \cr\cr
&\simeq & 8 \times 10^{-3} \left( {T_\fr \over m_{\rm WIMP}} \right)^{-5} \left({m_{\rm WIMP} \over M_{\rm pl}} \right)^{-5} 
\left( {\Gamma \over M_{\rm pl}} \right)^{2}  \left( {m_\sigma \over M_{\rm pl}} \right)^{1/2} \left( {\sigma_\ast \over M_{\rm pl}} \right)^{2}.
\label{eq:extra}
\end{eqnarray}
(This  can be obtained by using the expression for $T_{\rm dom}$ which is given later  in Eq.~\eqref{eq:T_dom} in Section \ref{subsub:newWIMP}.)


\bigskip
\bigskip
We expect the above expressions for $\Omega_{\rm WIMP}h^2$  to give only a rough estimate of the final DM abundance, within one order of magnitude. We have also computed  the precise prediction of  $\Omega_{\rm WIMP}$ by resorting to  a numerical computation, defining the ``correction" factor $A$ as $\Omega_{\rm WIMP}|_{\rm numerical} = A^{-1} \Omega_{\rm WIMP} |_{\rm analytical}$.
In the case where $Q_s \gg 1$
and $N_\dec > N_\fr > N_{\rm dom}$,
we have found that a correction factor $A \simeq 2.5$ is required.
We have also found that 
$A \simeq 1.8$ for  the case where $Q_s \gg 1$
and $N_\fr < N_{\rm dom}$ and
$A \simeq 1.1$ for $Q_s \simeq 1$. 

In Ref.~\cite{Salati:2002md}, a  more precise analytic formula for $\Omega_{\rm WIMP}$ has been proposed by using a different definition of the freeze-out time. We have extended this method  to apply it to  cases for which late time entropy production is significant and obtained an analytic formula for $\Omega_{\rm WIMP}$  which is in better agreement with the numerical results. In particular, in the case where $Q_s \gg 1$ and $N_\dec > N_\fr > N_{\rm dom}$, the correction factor is reduced to $A \simeq 1.7$. The derivation of this analytic formula is given in Appendix~\ref{sec:Salati}.

In Fig.~\ref{fig:iso_const}, the blue dashed line  corresponds to the isocurvature constraint obtained by naively requiring that 
the freeze-out occurs before the curvaton decay, i.e., $N_\fr > N_\dec$,  as usually assumed in  former works. 
To determine the region where $N_\fr > N_\dec$ is satisfied, we have numerically 
calculated the background evolutions of $\rho_r$ and $\rho_\sigma$. 
By contrast,  the red region corresponds to  the constraint derived by numerically evaluating $S_c$,
which automatically includes the effects of the curvaton-generated radiation. 
For both traditional and revised constraints, we resort to numerical calculation, even if  some rough analytic 
estimates can be derived, as we discuss below.

\subsubsection{Conventional constraint}

We have evaluated $N_\fr$ and $N_\dec$ by numerically solving the evolution of the system but it is also instructive to understand  the constraint via an analytic estimate. 
The  constraint derived from $N_\dec < N_\fr$  can be expressed as 
\begin{eqnarray}
\label{eq:T_fr_T_dec}
{T_{\rm fr} \over T_{\rm dec}} < 1.
\end{eqnarray}
In order to realize $\Omega_{\rm WIMP} \, h^2 = 0.12$  the freeze out temperature is such that 
$m_{\rm WIMP} / T_{\rm fr} \simeq 20$ (with our choice $m_{\rm WIMP} = 1$ TeV).
On the other hand, $T_{\rm dec}$ is determined by $\Gamma$  as
\begin{eqnarray}
{g_* \pi^2 \over 30} T_{\rm dec}^4 = M_{\rm pl}^2 \Gamma^2,
\end{eqnarray}
where  the decay time is defined by $\Gamma = H $. 
Therefore the requirement of Eq.~\eqref{eq:T_fr_T_dec} leads to 
\begin{eqnarray}
{m_{\rm WIMP} \over 20} \lesssim  \left( {30 \over  \, g_* \pi^2} \right)^{1/4} M_{\rm pl}^{1/2} \, \Gamma^{1/2},
\end{eqnarray}
from which we obtain 
\begin{eqnarray}
{\Gamma \over M_{\rm pl}} \gtrsim  10^{-33} \left({m_{\rm WIMP} \over 1~{\rm TeV}} \right)^2.
\end{eqnarray}
As expected, no $\sigma_\ast$ dependence appears in the constraint. 
Going back to Fig.~\ref{fig:iso_const}, 
we can see that the above formula provides a good estimate for the conventional constraint indicated 
by the blue dashed line.

\subsubsection{New constraint}
\label{subsub:newWIMP}

In Fig.~\ref{fig:iso_const}, 
the red excluded region is obtained by requiring $|S_c/\zeta | > 0.1$ where $S_c/\zeta$ is calculated numerically 
and the effects of the curvaton-generated radiation are automatically included as mentioned above.
In comparison with  the previous region satisfying  $N_\fr > N_\dec$ (blue dashed line), the new allowed region is significantly larger, as a consequence of  taking the curvaton-generated radiation effect into account, particularly  when the curvaton dominates the Universe early and the curvaton-generated radiation then becomes  the main component of radiation.

To understand better this new constraint,  an analytical argument is helpful.  As shown in the previous section, when the curvaton-generated radiation becomes dominant in the radiation component, 
the isocurvature fluctuations are significantly suppressed and the constraint $|S_c / \zeta |<0.1$ can be easily 
satisfied. Therefore the allowed parameter region can be roughly estimated by 
\begin{eqnarray}
\label{eq:t_dom_t_fr}
N_{\rm dom} <  N_{\rm fr}.
\end{eqnarray}
Instead of Eq.~\eqref{eq:t_dom_t_fr}, we can also use the following equivalent relation to derive the constraint:
\begin{eqnarray}
\label{eq:t_dom_t_fr_T}
{T_{\rm fr} \over T_{\rm dom}} < 1~.
\end{eqnarray}
When the curvaton-generated radiation is the main component in radiation, the Universe is dominated by the curvaton.
In such case, the energy density of the curvaton-generated radiation scales as $\rho_{r\sigma} \propto a^{-3/2}$~\cite{Kolb:text}, then 
the cosmic temperature is determined by
\begin{eqnarray}
T^4 \propto \rho_{r \sigma} \propto a^{-3/2}.
\label{eq:tempdep}
\end{eqnarray}
Based on this relation, $T_{\rm dom}$ can be estimated as
\begin{eqnarray}
T_{\rm dom} = T_{\rm dec} \left( {a_{\rm dom} \over a_{\dec}} \right)^{-3/8}
&=& \left( {90 \over g_{\ast, \dec} \pi^2} \right)^{1/4} \left({\Gamma  M_{\rm pl}}\right)^{1/2} \left( {a_{\rm dom} \over a_{\rm dec}} \right)^{-3/8}.
\label{eq:Tdom}
\end{eqnarray}
By using the fact that there is no entropy production between $N_{\sigma \osc}$ and  $N_{\rm dom}$, we also get
\begin{eqnarray}
T_{\rm dom} = T_{\sigma \osc} \left({a_{\rm dom} \over a_{\sigma \osc}}\right)^{-1} = \left( {90 \over g_{\ast, {\sigma \osc}} \pi^2} \right)^{1/4} \left({m_\sigma  M_{\rm pl}}\right)^{1/2} \left({a_{\rm dom} \over a_{\sigma \osc}}\right)^{-1},
\end{eqnarray}
where $a_{\sigma \osc}$ is the scale factor at the onset of the curvaton oscillations.
In addition, 
provided the curvaton dominates the universe at the decay, 
$a_{\rm dec}/a_{\sigma \osc}$ can be derived 
as follows,
\begin{equation}
\frac{a_{\rm dec}}{a_{\sigma \osc}} = \bigg(\frac{\rho_\sigma(t_{\rm dec})}{\rho_\sigma(t_{\sigma \osc})} \bigg)^{-1/3} = \left( \frac{3 \Gamma^2 M_\pl^2}{\frac{1}{2}m_\sigma^2 \sigma_*^2} \right)^{-1/3} = \frac{1}{6^{1/3}} \bigg(\frac{\Gamma}{m_\sigma} \bigg)^{-2/3}  \bigg(\frac{\sigma_*}{M_\pl}\bigg)^{2/3}.
\end{equation}
From these equations,
we obtain
\begin{eqnarray}
\label{eq:T_dom}
{T_{\rm dom} \over m_{\rm WIMP}} \simeq 0.38 \times \left({m_{\rm WIMP} \over M_{\rm pl}} \right)^{-1} 
\left( {\Gamma \over M_{\rm pl}} \right)^{2/5}  \left( {m_\sigma \over M_{\rm pl}} \right)^{1/10} \left( {\sigma_\ast \over M_{\rm pl}} \right)^{2/5}.
 \end{eqnarray}
By using the above equations, 
we find that the isocurvature constraint can be written as 
 \begin{eqnarray}
{\Gamma \over M_{\rm pl}} \gtrsim  10^{-38} \times 
\left( {m_{\rm WIMP}/T_\fr \over 20}\right)^{-5/2}
\left( {m_{\rm WIMP} \over 1~{\rm TeV}} \right)^{5/2}
\left( {m_\sigma \over 10^6~{\rm GeV}} \right)^{-1/4} \left( {\sigma_* \over M_{\rm pl}}\right)^{-1}.
\label{eq:WIMPconst}
\end{eqnarray}

In fact, the numerical result in Fig.~\ref{fig:iso_const} shows that
the actual isocurvature constraint (red line) is slightly larger than the above analytic estimate.
The difference can be explained as follows.
As shown in Fig.~\ref{fig:isocurv} the isocurvature perturbations
are actually suppressed 
not exactly for $N_\fr - N_{\rm dom} \gtrsim 0$ but rather for  $N_\fr - N_{\rm dom} \gtrsim \gamma~ (\gamma \simeq 2)$.
By taking into account this correction factor, $\gamma$, the constraint (\ref{eq:t_dom_t_fr_T}) can be rewritten as
\begin{equation}
{T_{\rm dom} \over T_\fr } > e^{3\gamma /8},
\end{equation}
where we have used $T \propto a^{-3/8}$ for $N > N_{\rm dom}$.
As a consequence, the lower bound for $\Gamma$ given in Eq.~(\ref{eq:WIMPconst}) is multiplied by a factor $e^{15 \gamma / 16}$. 

Furthermore, for $\sigma_\ast / M_{\rm pl} \gtrsim 10^{-2}$,
the dependence of the lower bound for $\Gamma$ on $\sigma_\ast / M_{\rm pl}$ seems to become weaker than
$\propto  \left( {\sigma_* / M_{\rm pl}}\right)^{-1}$ shown in Eq.~(\ref{eq:WIMPconst}). 
This feature can be qualitatively understood as follows.
This  region in  parameter space corresponds to the case where the freeze-out occurs during
when the curvaton-generated radiation dominates in the radiation component,
and the constraint is basically expressed as $T_\fr  < T_{\rm dom}$.
In this case, the expression for $\Omega_{\rm WIMP}$ is given by Eq.~\eqref{eq:Omega3} and, with the help of Eq.~\eqref{eq:extra}, 
it can be written as 
\begin{eqnarray}
\Omega_{\rm WIMP} h^2   
&\simeq & 0.12 
 \left( {m_{\rm WIMP} /T_\fr \over 20} \right) \left( {3.3 \times 10^{9}~ {\rm GeV}^{-1}  \over \lambda M_{\rm pl}} \right) \cr\cr
 && \quad \times
 \,\left( 1 - \Omega_{\sigma, \fr}\right)^{-1/2} \left( {m_{\rm WIMP} /T_\fr \over 20} \right)^5 \left( {m_{\rm WIMP} \over 1\,{\rm TeV}} \right)^{-5}
 \left( {\Gamma \over 1.5 \times 10^{-33} \, M_{\rm pl}}\right)^{5/2} \,.
 \label{eq:Omega4}
\end{eqnarray}
Recall that larger $\sigma_\ast$ and smaller $\Gamma$ generate larger entropy production by the decay of the curvaton,
 that is, larger $Q_s$.
A larger entropy production thus tends to reduce the relic abundance $\Omega_{\rm WIMP}$, as indicated by the dependence of $\Gamma$ in Eq.~(\ref{eq:Omega4}). 
But, as already mentioned, in our calculation we fix the relic abundance at $\Omega_{\rm WIMP} h^2 \simeq 0.12$,
and hence for larger $Q_s$ (corresponding to smaller $\Gamma$)
we take  $\lambda$ smaller in order  to keep $\Omega_{\rm WIMP}$ unchanged, which makes the freeze-out epoch 
earlier (i.e., $m_{\rm WIMP}/T_\fr$ becomes smaller (roughly, $m_{\rm WIMP}/T_\fr \propto \ln \lambda$) and the right hand side in Eq. (\ref{eq:WIMPconst}) becomes larger).

From the non-Gaussianity constraint, the parameter space where the curvaton dominates the Universe when it decays is preferred. 
Importantly, in such a region, the curvaton-generated radiation also tends to be dominant in the radiation component 
and  the isocurvature fluctuations 
are significantly suppressed. 
This means that part of the parameter space which was believed to be excluded by the isocurvature constraints is in fact allowed.

\subsection{Axion DM}

\subsubsection{Abundance}

We now present our results for axionic DM,  taking into account the entropy production due to the curvaton decay and the 
effects of the curvaton-generated radiation.
The final abundance of the axion DM depends on the details of the scenario. One can distinguish  three cases as listed below.

\bigskip
\noindent
$\bullet$ {\bf Case 1 : axion oscillations begin during RD era (before/after the curvaton decay)} 

In the case where the axion starts to oscillate during radiation domination,  either before or after the curvaton decay, one has\footnote{
Conventionally, the beginning of the axion oscillation is defined by $3H=m(T)$ in the literature. In this paper, however, we adopt the same definition as Eq.~(\ref{eq:osc_begin}) which is used to calculate the isocurvature perturbation. This affects  the final abundance  by a factor of about 2.
}
\begin{equation}
H = m(T)~~\text{with}~~H = \bigg(\frac{\pi^2 g_*}{90} \bigg)^{1/2}\frac{T^2}{M_\pl}\,,
\end{equation}
which leads to
\begin{equation}
T_{\chi\osc} \simeq 1~{\rm GeV} \bigg(\frac{F_a}{10^{12}~{\rm GeV}}\bigg)^{-\frac{1}{\beta+2}}.
\label{TchioscRD}
\end{equation}
Here and in what follows, we set $g_{*\chi{\rm osc}} = 80$ as an indicative reference value.
The ratio between the axion number density and the entropy density at that time is 
\begin{equation}
\frac{n_\chi}{s} \bigg|_{\chi\osc} = \frac{T}{8m(T)}\bigg|_{\chi\osc} \bigg(\frac{\chi_i}{M_\pl}\bigg)^2.
\end{equation}
From now on, we set $\chi_i = F_a \, (\theta_i = 1)$ for simplicity.
$n_\chi$ is an adiabatic invariant quantity but if the curvaton decay occurs after the axion production, it is diluted by the entropy production due to the curvaton decay.
The resultant density parameter of the axion CDM is obtained by multiplying the entropy dilution factor, $Q_s^{-1}$, defined by Eq.~(\ref{def_Qs}), 
a correction factor $\kappa$ associated with the non-adiabaticity and the present axion mass and being divided by $\rho_{\rm crit}/s_0 h^2$ 
\begin{equation}
\Omega_\chi h^2 \simeq 0.02 \kappa \bigg(\frac{F_a}{10^{12}~{\rm GeV}} \bigg)^{\frac{\beta+3}{\beta+2}} 
\times \begin{cases} 1 ~~&\text{for}~~T_{\chi\osc} < T_{\rm dec} \\ Q_s^{-1} ~~&\text{for}~~T_{\chi\osc} > T_{\rm dec} \end{cases}.
\label{Omega_chi_RD}
\end{equation}
We have estimated the correction factor by taking the ratio between analytic and numerical results of the final axion abundance and found $\kappa \sim 5$ in this case.
In the case where  the axion starts to oscillate when the fraction of the energy density of the curvaton becomes large, $\kappa$ can be as large as 8.

\bigskip
\noindent
$\bullet$ {\bf Case 2 : axion oscillation begins during $\sigma$D era and $T_{\chi\osc} > T_{\rm dom}$} 

If the axion oscillations begin in the curvaton dominated era, but while  the radiation component is still  dominated by the inflaton-generated radiation,  
the Hubble parameter at the onset of axion oscillations is given by
\begin{equation}
H_{\chi\osc} = \frac{\rho_\sigma^{1/2}(t_{\chi\osc})}{\sqrt{3}M_\pl} = \frac{\rho_\sigma^{1/2}(t_{\osc})}{\sqrt{3}M_\pl} \bigg(\frac{T_{\chi\osc}}{T_\osc} \bigg)^{3/2},
\end{equation}
and we obtain
\begin{equation}
T_{\chi\osc} = 0.08~{\rm GeV} \bigg(\frac{m_\sigma}{10^6~{\rm GeV}} \bigg)^{-\frac{1}{4(\beta+3/2)}} \bigg(\frac{\sigma_*}{M_\pl}\bigg)^{-\frac{1}{\beta+3/2}} \bigg(\frac{F_a}{10^{12}~{\rm GeV}} \bigg)^{-\frac{1}{\beta+3/2}}\,.
\end{equation}
The initial energy density of the axion is related to that of the curvaton through
\begin{equation}
\frac{\rho_\chi}{\rho_\sigma} \bigg|_{\chi\osc} = \frac{\frac{1}{2} m(T)^2 F_a^2}{3 H^2 M_\pl^2} \bigg|_{\chi\osc} =  \frac{1}{6} \bigg(\frac{F_a}{M_\pl}\bigg)^2,
\end{equation}
and by using the conservation of $\rho_\sigma/s$, one obtains
\begin{equation}
\frac{\rho_\chi}{s}\bigg|_{\chi\osc} = \frac{\rho_\chi}{\rho_\sigma} \bigg|_{\chi\osc} \frac{\rho_\sigma}{s} \bigg|_{\chi\osc} 
= \frac{1}{6} \bigg(\frac{F_a}{M_\pl} \bigg)^2 \frac{\rho_\sigma}{s} \bigg|_{\osc} \frac{\mathcal{S}_i}{\mathcal{S}_{\chi \osc}} = \frac{1}{48} T_\osc \bigg(\frac{\sigma_*}{M_\pl}\bigg)^2 \bigg(\frac{F_a}{M_\pl}\bigg)^2\frac{\mathcal{S}_i}{\mathcal{S}_{\chi \osc}}.
\end{equation}
Because the curvaton dominates the Universe at the curvaton decay, a significant amount of entropy is produced. 
Hence the present value of $\rho_\chi/s$ can be given by
\begin{equation}
\frac{\rho_\chi}{s}\bigg|_{0} = \frac{\mathcal{S}_{\chi\osc}}{\mathcal{S}_f} \frac{\kappa m_*}{m(T_{\chi\osc})}  \frac{\rho_\chi}{s}\bigg|_{\chi\osc}  = \frac{\kappa T_{\rm dec} m_*}{8m(T_{\chi\osc})} \bigg(\frac{F_a}{M_\pl} \bigg)^2,
\label{rho_chi_s_0}
\end{equation}
where  we have used Eq.~\eqref{eq:Q_s} in the last equality.
Finally, one can obtain the present density parameter of the axion DM,
\begin{equation}
\Omega_\chi h^2 = 3 \times 10^{-7} \kappa \bigg(\frac{m_\sigma}{10^6~{\rm GeV}}\bigg)^{-\frac{\beta}{4\beta+6}} \bigg(\frac{\sigma_*}{M_\pl} \bigg)^{-\frac{\beta}{\beta+3/2}} \bigg(\frac{T_{\rm dec}}{0.1~{\rm GeV}} \bigg) \bigg(\frac{F_a}{10^{12}~{\rm GeV}}\bigg)^{\frac{\beta+3}{\beta+3/2}}.
\end{equation}
By comparing with our numerical results, we have found $\kappa \sim 3.5$. However  $\kappa$ can be as large as 5 near  the boundary between cases 1 and 2.

\bigskip
\noindent
$\bullet$ {\bf Case 3 : axion oscillations begin during $\sigma$D era and $T_{\chi\osc} < T_{\rm dom}$} 

In this case, $T_{\chi\osc}$ is determined by the temperature of the curvaton-generated radiation,
\begin{equation}
T =\bigg( \frac{10g_{*\rm dec}}{\pi^2g_*^{2}}\bigg)^{1/8} (3H T_{\rm dec}^2 M_\pl)^{1/4},
\end{equation}
where $H$ is to be replaced with $m(T)$. Then the temperature at the onset of axion oscillation is
\begin{equation}
T_{\chi\osc} = 0.5~{\rm GeV}  \bigg(\frac{T_{\rm dec}}{0.1~{\rm GeV}} \bigg)^{\frac{2}{\beta+4}} \bigg(\frac{F_a}{10^{12}~{\rm GeV}} \bigg)^{-\frac{1}{\beta+4}},
\label{Tchiosc_sD2}
\end{equation}
where we set $g_{*\rm dec} = 10$ and $g_{*\rm dom}=80$.
Substituting it into Eq.~(\ref{rho_chi_s_0}), the density parameter of the axion CDM is found to be given by
\begin{equation}
\Omega_\chi h^2 = 1 \times 10^{-4} \kappa \bigg(\frac{T_{\rm dec}}{0.1~{\rm GeV}}\bigg)^{\frac{3\beta+4}{\beta+4}} \bigg(\frac{F_a}{10^{12}~{\rm GeV}} \bigg)^{\frac{\beta+8}{\beta+4}},
\label{Omega_chi_sD2}
\end{equation}
 where $\kappa \sim 1.5$ in this case.

\subsubsection{Conventional constraint}

According to the conventional argument, the axion should start to oscillate after the curvaton decay in order to be consistent with the present  upper bound on isocurvature perturbations.
This corresponds to the constraint  $N_{\chi\osc} > N_{\rm dec}$ or $T_{\chi\osc} < T_{\rm dec}$.
Because the universe is dominated by radiation after the curvaton decay, one can use Eq.~(\ref{TchioscRD}) for $T_{\chi\osc}$.
Note that there is no entropy production after the onset of the axion oscillations and $F_a$ can be fixed by using Eq.~(\ref{Omega_chi_RD}) if we assume $\Omega_\chi = \Omega_{\rm DM}$.
Then, $T_{\chi\osc} < T_{\rm dec}$ leads to
\begin{equation}
\frac{\Gamma}{M_{\rm pl}} \gtrsim \frac{m_a(T_{\chi\osc})}{M_{\rm pl}} \simeq 10^{-36},
\end{equation}
thus providing an upper bound on the decay rate of the curvaton, which corresponds to the dashed blue horizontal line in Fig.~\ref{fig:iso_const_axion}.

\subsubsection{New constraint}

\begin{figure}[tp]
\centering
\includegraphics [width = 10cm, clip]{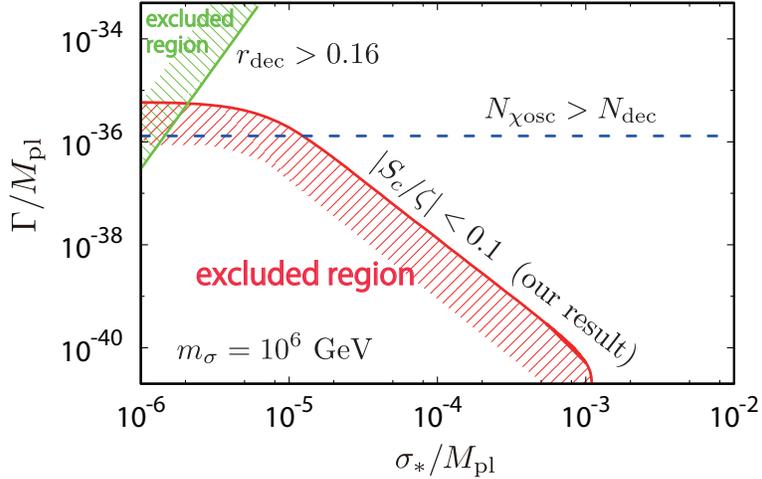}
\caption{
Isocurvature and non-Gaussianity constraints in the $\Gamma$--$\sigma_\ast$ plane for  axion DM, obtained  numerically.
The red curve is our new constraint and the blue horizontal line is the conventional constraint.
The green line shows the constraint from $f_{\rm NL}$.
We have taken $m_\sigma = 10^6$~GeV and 
$\beta=3.34$ (corresponding to the QCD axion case). The allowed region is the top right part of the figure.
}
\label{fig:iso_const_axion}
\end{figure}

We have shown in the previous section that the isocurvature constraint is milder than previously thought and  is roughly given by $T_{\rm dom} \gtrsim T_{\chi\osc}$.
In this situation, one can use Eq.~(\ref{Tchiosc_sD2}) for $T_{\chi\osc}$.
$F_a$ can be fixed by using Eq.~(\ref{Omega_chi_sD2}) with $\Omega_\chi = \Omega_{\rm DM}$ and $T_{\rm dom} \gtrsim T_{\chi\osc}$ leads to 
\begin{equation}
\frac{\Gamma}{M_\pl} \gtrsim 1 \times 10^{-50} \times \kappa^{\frac{10}{4\beta + 7}} \bigg(\frac{m_\sigma}{10^6~{\rm GeV}} \bigg)^{-\frac{\beta+8}{4\beta+7}} \bigg(\frac{\sigma_*}{M_\pl}\bigg)^{-\frac{4(\beta+8)}{4\beta+7}}.
\label{eq:iso_const_axion}
\end{equation}
Fig.~\ref{fig:iso_const_axion} shows numerical result of the excluded region in $\Gamma$--$\sigma_*$ plane. 
The region below the red curve is excluded by  the  CDM isocurvature constraint  $|S_c/\zeta|<0.1$, which translates into  the inequality (\ref{eq:iso_const_axion}). The dashed blue horizontal line corresponds to the conventional upper bound, discussed earlier,  and the green line represents the constraint from non-Gaussianity. 
Similarly to the WIMP case, one can also take into account the correction factor $\gamma \equiv N_{\chi\osc}-N_{\rm dom}$, which entails a multiplication of the right hand side in Eq.~(\ref{eq:iso_const_axion})  by $e^{\frac{15(\beta+8)}{4(4\beta+7)}\gamma}$.
Comparing the analytic formula with the numerical result, we have found $\gamma \simeq 3$.
Finally, one notes that the red line drops sharply  near $\sigma_*/M_{\rm pl} \sim 10^{-3}$. This is due to the disappearance of the temperature dependence of the axion mass (see Eq.~(\ref{eq:m_chi})).

\section{Conclusion}

In this paper, we have made a detailed investigation of DM isocurvature fluctuations in the curvaton scenario, when DM is composed of 
 WIMPs or axions, and revisited the constraints that can be inferred from the non-observation of isocurvature perturbations in the CMB. 
In previous works, it is usually considered that  isocurvature fluctuations are too large to be consistent with observations as soon as  DM is created before the curvaton decay. 

However, in the present work, we have found that more  scenarios are still acceptable. Indeed, one can still include models
 where DM is  created before the completion of the curvaton decay, provided the DM freeze-out (for WIMPs, or the onset of oscillations for axions) occurs  when the curvaton-generated radiation dominates 
the radiation component. In this situation, the isocurvature perturbations are also suppressed. 
Interestingly,  curvaton scenarios where the curvaton tends to dominate the energy content of the Universe over a long period of time are favoured by the current constraints on primordial non-Gaussianity. 

In summary,  our results show that the usual constraints  on curvaton models from isocurvature fluctuations can be partially relaxed, thus opening a new window for viable scenarios. In this paper, we have only investigated the  isocurvature fluctuations associated with  DM.  Similar argument should also hold for baryonic isocurvature perturbations, which could lead to useful consequences for  models of baryogenesis in the framework of the curvaton scenario. 
Furthermore, it would be interesting to study other type of CDM such as feebly interacting massive particle (FIMP)  \cite{Hall:2009bx}, 
strongly interacting massive particle (SIMP) \cite{Hochberg:2014dra,Hochberg:2014kqa}.

\section*{Acknowledgments}

T.T would like to thank APC for its hospitality during his visit, where part of this work has been done.  
This work  is partially supported by JSPS KAKENHI Nos.~15H05888 (TT, SY), 15K05084 (TT), 15K17659 (SY)
and 16H01103 (SY). 
N.K. acknowledges the support by Grant-in-Aid for JSPS Fellows and the Max-Planck-Gesellschaft, the Korea Ministry of Education, Science and Technology, Gyeongsangbuk-Do and 
Pohang City for the support of the Independent Junior Research Group at the Asia Pacific Center for Theoretical Physics.

\appendix 

%
\bigskip
\bigskip
\bigskip

\section{Formalism with curvaton-generated radiation}
\label{sec:dillute_plasma}

As mentioned in Section~\ref{sec:ITF},  it has been usually considered that 
sizable isocurvature fluctuations are generated when WIMP DM particles  freeze out before the curvaton decay, or 
a scalar field such as the axion begins to oscillate before the curvaton decay. 
However,  if the curvaton-generated radiation dominates over that produced by the inflaton sector, 
the isocurvature fluctuations are drastically suppressed. We have demonstrated this via  numerical calculations as 
well as semi-analytic derivations in Section~\ref{sec:ITF}. 

In this appendix, we show that   the suppression of isocurvature fluctuations can also be recovered via a fully analytic derivation based on the (somewhat artificial) separation of the total radiation energy density $\rho_r$ into two components, 
\begin{equation}
\label{eq:rho_r_tot_dilute}
\rho_r = \rho_{r\phi} + \rho_{r\sigma}, 
\end{equation}
where $\rho_{r\phi}$ and $\rho_{r\sigma}$ are the energy densities of the radiations produced by the inflaton  and  by the curvaton, respectively. 
Notice that  the scaling law of the curvaton-generated radiation is different from the usual radiation and it can be written as  \cite{Kolb:text}
\begin{equation}
\rho_{r\sigma} \propto a^{-p},
\end{equation}
with $p= 3/2$  during the curvaton-dominated (matter-dominated) epoch and $p=1$  during the radiation-dominated epoch. 
With the above parametrization, the equation of state for the curvaton-generated radiation can be written as $1+w_{r\sigma} = p/3$ 
and its perturbation is given by 
\begin{equation}
\zeta_{r\sigma} = \delta N + \frac{1}{p} \ln \left( \frac{\rho_{r\sigma} (t, \vec{x})}{\bar{\rho}_{r\sigma}(t)} \right).
\end{equation}
From the above equation, one can write its energy density as
\begin{equation}
\rho_{r\sigma} (t, \vec{x}) = \bar{\rho}_{r\sigma} (t) e^{p (\zeta_{r\sigma} - \delta N)}.
\end{equation}
As for the inflaton-generated radiation, its energy density scales as $\rho_{r\phi} \propto a^{-4}$, which implies $\rho_{r\phi} (t,\vec{x}) = \bar{\rho}_{r\phi} (t) e^{4 (\zeta_{r\phi} - \delta N)}$.

Taking into account the two radiation components, the temperature is given by 
\begin{equation}
\label{eq:temp_mix_2}
T \equiv \left[ \frac{30}{\pi^2 g_\ast}  \left( \rho_{r\phi}  + \rho_{r \sigma} \right) \right]^{1/4},\
\end{equation}
and  the temperature fluctuations are related to the radiation perturbations $\zeta_{r\phi}$ and $\zeta_{r\sigma}$ according to the expression
\begin{equation}
\label{eq:delta_T_dilute}
\delta_T = (1- f) (\zeta_{r\phi} - \delta N) + \frac{p}{4} f ( \zeta_{r\sigma} - \delta N),
\end{equation}
where we have introduced the fraction of the curvaton-generated radiation in the total radiation component,
\begin{equation}
f \equiv \frac{\rho_{r\sigma}}{\rho_r} =  \frac{\rho_{r\sigma}}{\rho_{r\phi} + \rho_{r\sigma}}\,.
\end{equation}
Evaluating $\delta_T$  at the transition time and using  Eqs.~\eqref{eq:delta_Gamma_delta_T} and \eqref{eq:delta_T_dilute}, we obtain
\begin{equation}
\label{eq:delta_N_H_dilute}
\delta N_{\cal H} = \frac{1}{3 \tilde{\Omega} - 2 \alpha (1-f + fp /4 )} \left(  3 \sum_i \tilde{\Omega}_{i} \, \zeta_{i} - 2\alpha (1-f) \zeta_{r\phi} - \frac{1}{2} \alpha f p\,  \zeta_{r\sigma} \right).
\end{equation}
When $f=0$ (i.e., no curvaton-generated radiation), the above expression reduces to Eq.~\eqref{eq:deltaN_H}.
Since we separate the radiation component into the curvaton- and inflaton-generated ones, $ \tilde{\Omega}$ and $  \sum_i \tilde{\Omega}_{i} \zeta_{i}$ are
written as 
\begin{eqnarray}
3\tilde{\Omega} &=& 4\Omega_{r\phi} + 3 \Omega_\sigma + p\, \Omega_{r\sigma}, \\
3 \sum_i \tilde{\Omega}_{i}\,  \zeta_{i} &=& 4\Omega_{r\phi}  \zeta_{r\phi} + 3 \Omega_\sigma \zeta_\sigma + p \, \Omega_{r\sigma} \zeta_{r\sigma}.
\end{eqnarray}
Furthermore, by equating $\rho_{r\sigma}$ and $\rho_\sigma$ at the time of the curvaton decay as\footnote{
This relation can be considered to determine the normalization of the energy density of the curvaton-generated radiation.
}
\begin{equation}
\bar{\rho}_{r\sigma} e^{p(\zeta_{r\sigma} - \delta N_{\rm dec})}  =
\bar{\rho}_{\sigma} e^{3(\zeta_\sigma - \delta N_{\rm dec})},
\end{equation}
at linear order, we can relate $\zeta_{r\sigma}$ and $\zeta_\sigma$ as follows:
\begin{equation}
p \, \zeta_{r\sigma} = \left[ 3 + (p-3) r_{\rm dec} \right] \zeta_\sigma,
\end{equation}
where we have used the fact that $\delta N_{\rm dec} = r_{\rm dec} \zeta_\sigma$ with 
$r_{\rm dec}$ being given by 
\begin{equation}
\label{eq:r_dec_separate}
r_{\rm dec} = \left. \frac{3 \Omega_\sigma + 3\Omega_{r\sigma}}{4 \Omega_{r\phi} + 3 \Omega_\sigma + 3 \Omega_{r\sigma}} \right|_{\rm dec},
\end{equation}
which is a generalization of the one given in Eq.~\eqref{eq:r_dec}.

We now consider  the CDM perturbation $\zeta_c$. From Eqs.~\eqref{eq:n_c} and \eqref{eq:delta_n}, we can write 
\begin{equation}
\zeta_c = \frac13 \left( \nu \delta_T + 3 \delta N_{\cal H} \right).
\end{equation}
In the simplest curvaton scenarios, the inflaton fluctuations are negligible compared to those of the curvaton and we thus
 set  $\zeta_{r\phi}=0$. 
Substituting the expressions for $\delta_T$ and $\delta N_{\cal H}$ given in Eqs.~\eqref{eq:delta_T_dilute} and \eqref{eq:delta_N_H_dilute}, we finally obtain
\begin{equation}
\label{eq:zeta_c_dilute}
\zeta_c = \frac13 \left[
\frac{\nu f}{4} \left( 3 + (p-3) r_{\rm dec} \right) 
+ \left\{ 3 - \nu \left( 1 - f + \frac{fp}{4} \right) \right\}
\left( 
\frac{ 3 \Omega_\sigma + \left( \Omega_{r\sigma} - \alpha f/2 \right) \left( 3 + (p-3) r_{\rm dec} \right) }
{4 -\Omega_\sigma + (p-4) \Omega_{r\sigma} -2 \alpha \left( 1 - f + fp/4 \right) }
\right)
\right]\zeta_\sigma.
\end{equation}
Notice that the above expression is a general formula where we just assumed $\Gamma = \Gamma (T)$, 
and all quantities on the right hand side, except $r_{\rm dec}$, should be evaluated at the time of the transition.
For the WIMP case, we can just set 
\begin{equation}
\alpha = \frac{m}{T} + \frac32 + \frac{q}{2},
\qquad
\nu = \frac{m}{T} + \frac32,
\end{equation}
as already given in Eqs.~\eqref{eq:alpha_wimp} and \eqref{eq:nu_wimp}. 
For the axion case, as given in Eq.~\eqref{eq:alpha_nu_axion}, we can use
\begin{equation}
\alpha = \nu = -\beta.
\end{equation}

From Eq.~\eqref{eq:zeta_c_dilute}, we can easily see why the isocurvature fluctuations vanish even if CDM particles freeze out before the curvaton 
decay, or the axion begins to oscillate before the curvaton decay when the curvaton-generated radiation dominates in the radiation component. 
In such cases, the radiation energy density is $\rho_r \simeq \rho_{r\sigma}$, which corresponds to $f \simeq 1$ and 
$\Omega_{r\phi} \ll \Omega_{r\sigma}, \Omega_\sigma$. Therefore, by evaluating $\zeta_c$ using Eq.~\eqref{eq:zeta_c_dilute} with
\begin{equation}
f \simeq 1,
\qquad
r_{\rm dec} =1, 
\qquad
\Omega_\sigma + \Omega_{r\sigma} \simeq 1,
\end{equation}
we obtain $\zeta_c \simeq \zeta_\sigma$, which means that $S_c \simeq 0$. Hence the isocurvature perturbation is suppressed even if the transition,  such as the WIMP freeze-out or the onset of axion oscillations of the axion, occurs before the curvaton decay, as confirmed by the numerical treatment discussed in the main text.

\section{Another analytic expression for $\Omega_{\rm WIMP}$}
\label{sec:Salati}

The goal of this appendix is to derive a more precise estimate of the final $\Omega_{\rm WIMP}$, inspired by Ref. \cite{Salati:2002md}.

Following Ref. \cite{Salati:2002md},
let us introduce a variable $\f$ defined as
\begin{eqnarray}
\f \equiv {n_{\rm WIMP} \, a^3 \over a_{\bar{\rm fr}}^3 T_{\bar{\rm fr}}^3},
 \end{eqnarray}
where a subscript ``$\bar{\rm fr}$" denotes
a time of the WIMP freeze-out defined in the following (see Eq.~\eqref{eq:freeze_appB}). 
It should be noticed that the definition of the freeze-time in this appendix  is different from 
the one adopted in the main text.

 By using this variable, the evolution equation for $n_{\rm WIMP}$ can be written as
\begin{eqnarray}
{d \f \over dt} + \lambda n_{\rm WIMP}\,  \f = \lambda {a_{\bar{\rm fr}}^3 T_{\bar{\rm fr}}^3 \over a^3} (\f^{({\rm eq})})^2.
\label{eq:feq}
\end{eqnarray}
In the above equation, we can identify two time scales, $\tau_{\rm rel}$ and $\tau_{\rm eq}$,
 respectively defined by
\begin{eqnarray}
&& \tau^{-1}_{\rm rel} = \lambda\,  n_{\rm WIMP}, \cr\cr
&& \tau^{-1}_{\rm eq} = - {d \over dt} \ln  \left[ (\f^{({\rm eq})})^2 a^{-3}\right]
= - 3 H - 2 \left( {3 \over 2} + {m_{\rm WIMP} \over T}\right) {d \ln T \over dt}\,,
\end{eqnarray}
and the freeze-out  corresponds to the instant when these two time scales coincide, i.e. 
\begin{eqnarray}
\label{eq:freeze_appB}
\tau_{\rm rel}( t_{\bar{\rm fr}}) = \tau_{\rm eq}( t_{\bar{\rm fr}}).
\end{eqnarray}
After freeze-out, the right hand side of Eq. (\ref{eq:feq}) becomes negligible,
the equation for $\f$ reduces to
\begin{eqnarray} 
{d\f \over dt} + \lambda {a_{\bar{\rm fr}}^3 T_{\bar{\rm fr}}^3 \over a^3} \f^2 \simeq 0\,.
\end{eqnarray}
This equation means that $\f$ slightly evolves in time even after the freeze-out time,
and by integrating this equation, we obtain
\begin{eqnarray}
\f (t) = {\f_{\bar{\rm fr}} \over 1 + \mu (t)},
\end{eqnarray}
where we have introduced an evolution factor $\mu (t)$,  defined by
\begin{eqnarray}
\mu (t) = \lambda n_{\rm WIMP,\bar{fr}} \int^t_{t_{\bar{\rm fr}}} e^{-3 (N - N_{\bar{\rm fr}})} dt.
\end{eqnarray}
Using the above expression, $\Omega_{\rm WIMP}$ is given by
\begin{eqnarray}
\Omega_{\rm WIMP} h^2 &=& 
\frac{m_{\rm WIMP} n_{\rm WIMP,0} }{\rho_{\rm crit} /h^2} 
=
\frac{m_{\rm WIMP} }{\rho_{\rm crit} /h^2} \left( \frac{a_c}{a_0}\right)^3  n_{\rm WIMP, c} \notag \\
&=&
m_{\rm WIMP}  \frac{s_0}{\rho_{\rm crit} /h^2} \left( \frac{a_c^3 s_c}{a_0^3 s_0}\right)  \frac{n_{\rm WIMP, c}}{s_c} \notag \\
 &\simeq &  
 0.12   \left( {m_{\rm WIMP} /T_\fr \over 20} \right) \left( {3.3 \times 10^{9}~ {\rm GeV}^{-1}  \over \lambda M_{\rm pl}} \right) 
\left( 1 - \Omega_{\sigma, \fr}\right)^{-1/2} \frac{1}{1+ \mu_0} \notag \\
&&
\times \left[ -3 - 2 \left( \displaystyle{3 \over 2} + {m_{\rm WIMP} \over T_{\bar{\rm fr}} } \right)  
\left. \displaystyle{d \ln T \over dN}  \right|_{t = t_{\bar{\rm fr}}} \right]
\times
\begin{cases} 
1 ~~&\text{for}~~T_{c} < T_{\rm dec} \\ R_s^{-1} ~~&\text{for}~~T_{c} > T_{\rm dec} 
\end{cases},
\end{eqnarray}
where $\mu_0$ is evaluated at $t = t_0$ and we have assumed $T_\fr \simeq T_c$.

{}

\end{document}